\documentclass{article}

\usepackage{microtype}
\usepackage{graphicx}
\usepackage{subcaption}
\usepackage{booktabs} % for professional tables

\usepackage{hyperref}

\usepackage[accepted]{icml2026}

\usepackage{amsmath}
\usepackage{amssymb}
\usepackage{mathtools}
\usepackage{amsthm}

\usepackage[capitalize,noabbrev]{cleveref}

\theoremstyle{plain}

\theoremstyle{definition}

\theoremstyle{remark}

\usepackage[textsize=tiny]{todonotes}

\icmltitlerunning{Extracting Recurring Vulnerabilities from Black-Box LLM-Generated Software}

\begin{document}

\twocolumn[
  \icmltitle{Extracting Recurring Vulnerabilities from Black-Box LLM-Generated Software}



  \icmlsetsymbol{equal}{*}

% Corresponding
% Code <>; 
% <>; 
% <>; Amit LeVi – Main
% Advisor <>; 
% <>.

  \begin{icmlauthorlist}
    \icmlauthor{Tomer Kordonsky}{equal,tech}
    \icmlauthor{Amit Levi}{equal,tech}
    \icmlauthor{Maayan Yamin}{equal,tech}
    \icmlauthor{Noam Benzimra}{equal,tech}
    \icmlauthor{Avi Mendelson}{tech}
  \end{icmlauthorlist}

  \icmlaffiliation{tech}{Technion - Israel Institute of Technology}

  \icmlcorrespondingauthor{Tomer Kordonsky}{tkordonsky@campus.technion.ac.il}
  \icmlcorrespondingauthor{Amit Levi}{amitlevi@campus.technion.ac.il}
  \icmlcorrespondingauthor{Maayan Yamin}{maayan.yamin@campus.technion.ac.il}
  \icmlcorrespondingauthor{Noam Benzimra}{noambenzimra@campus.technion.ac.il}
  \icmlcorrespondingauthor{Avi Mendelson}{mendlson@technion.ac.il}

  \icmlkeywords{Machine Learning, ICML}

  \vskip 0.3in
]



\printAffiliationsAndNotice{}  % no special notice (required even if empty)

\begin{abstract}
Large language models are increasingly deployed as core engines for automated code generation, accelerating software development while also emitting insecure programs. Existing defenses rely mainly on post-hoc scanning and treat each sample in isolation, leaving open a more predictive question: are these failures recurring enough that hidden vulnerabilities can be inferred from a model's visible outputs alone? We study this threat model, which we call \emph{vulnerability persistence}, and introduce the Feature-Security Table (FSTab), a model-specific mapping from observable frontend features to recurrent backend vulnerabilities constructed from generated applications labeled with static-analysis findings. FSTab is queried using only public UI actions and endpoints, without source-code access, and across six code LLMs and five WebGenBench categories it achieves perfect attack success and coverage in multiple held-out categories while remaining strong under cross-category transfer, reaching up to 85.67\% average attack success and 84.97\% average coverage when the target category is excluded during construction. We further introduce a model-centric recurrence suite over features, prompt rephrasings, and application categories, and find that cross-category transfer exceeds within-category recurrence, suggesting that many insecure implementations reflect stable model-level coding habits rather than isolated prompt artifacts. These results expose a predictive black-box attack surface in LLM-generated software, motivate evaluation of recurring model behaviors, and our code is available at \href{https://anonymous.4open.science/r/FSTab-024E/README.md}{\includegraphics[height=1em]{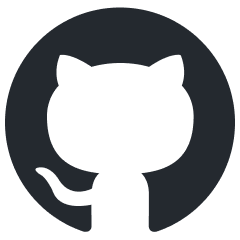}}.
\end{abstract}

\begin{figure}[ht]
    \centering
    \includegraphics[width=\linewidth]{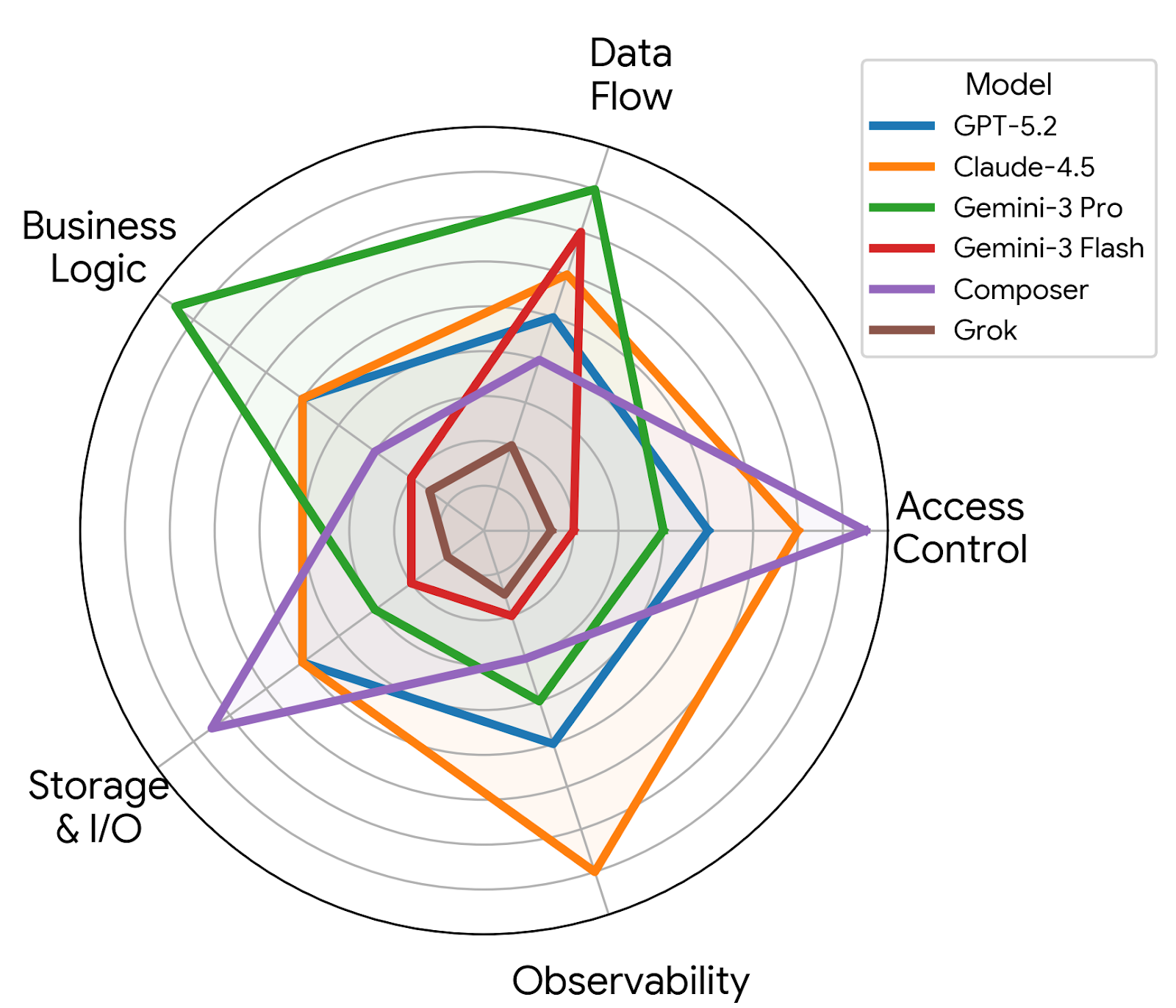}
    \caption{Architectural Vulnerability Fingerprints.}
    \label{fig:radar_fingerprints}
\end{figure}

\section{Introduction}
\label{sec:intro}

Large language models are now embedded in IDEs and agentic development workflows, where they synthesize, edit, and refactor production code at increasing scale \cite{chen2021evaluating,li2022competition,nijkamp2022codegen,fried2022incoder,wang2021codet5,guo2022unixcoder,li2023starcoder,roziere2023code}. This shift improves developer productivity, but it also imports security risk: generated programs frequently contain high-impact vulnerabilities, and current defenses still rely primarily on post-hoc static analysis or benchmark-style testing after the code has already been produced \cite{pearce2025asleep,siddiq2022securityeval,tony2023llmseceval,bhatt2023purple,wang2024your,li2025safegenbench,dubniczky2025castle}.

This deployment model changes what an attacker can realistically observe. In many practical settings, the application is public while the source repository, prompting history, and model logs are not. A black-box adversary can still browse the UI, trigger endpoints, inspect network behavior, and enumerate which workflows are present. If a model repeatedly implements those workflows with the same insecure backend templates, then the visible surface of a generated application leaks useful information about hidden vulnerabilities even when the code remains inaccessible.

Our starting observation is that code LLMs do not fail independently across tasks. They often reuse architectural templates, endpoint patterns, and implementation idioms. If those recurring templates are insecure, then hidden backend vulnerabilities may be partly predictable from the visible surface of an application. Public workflows such as login, search, file upload, or payment can therefore leak information about likely backend flaws even when the source code is unavailable. We study this threat model as \emph{vulnerability persistence}.

The key question is therefore not only whether a model can emit vulnerable code, but whether it does so with enough regularity that an adversary can build a reusable prior. A repeated coupling such as ``login: insecure randomness'' or ``upload: unsafe file handling'' turns isolated bugs into a model-level attack surface. This perspective differs from standard per-sample benchmarking, which usually scores each generated program independently and discards the structure shared across outputs.

Prior work has documented insecure code generation and, separately, shown that black-box LLM outputs can expose stable signatures useful for extraction or fingerprinting \cite{tramer2016stealing,shokri2017membership,carlini2021extracting}. We connect these lines of research: rather than asking only whether a generated program is vulnerable, we ask whether feature-conditioned vulnerability patterns are stable enough to support black-box inference of hidden vulnerabilities across programs, prompts, and domains.

We introduce the \emph{Feature--Security Table} (FSTab), a model-specific lookup table that maps observable frontend features to recurring backend vulnerabilities. FSTab is built from previously generated applications labeled with static-analysis findings and can then be queried against a deployed target using only visible UI actions and endpoints. The workflow in Figure~\ref{fig:e2e_main} makes this operational: generate a profiling corpus, extract visible features and scanner findings, aggregate a compact feature-vulnerability database, and use it to prioritize likely hidden flaws in unseen targets.

\begin{figure*}[ht]
    \centering
    \includegraphics[width=\textwidth]{imgs/End_to_End.pdf}
    \caption{Overview of the FSTab workflow. For each code model, we build a feature-security lookup table from generated applications and then query that table using only observable target functionality.}
    \label{fig:e2e_main}
\end{figure*}

Across six SOTA code LLMs and five WebGenBench domains \cite{lu2026webgen}, we find that this simple procedure often recovers a compact set of likely hidden vulnerabilities, including in cross-domain settings where the target domain is excluded during construction. This makes the paper's central claim concrete: model-level recurrence is not merely an analysis artifact, but a usable attack prior for black-box security auditing. The online attack input is only the set of visible user-facing actions exposed by the target application; source-level feature extraction is used only offline on attacker-owned generations to standardize the construction corpus at scale, not because victim-side source access is required.

Our contributions are threefold:
\begin{itemize}
    \item \textbf{A practical black-box attack.} We present FSTab, a feature-conditioned lookup attack that prioritizes hidden backend vulnerabilities from only observable application functionality.
    \item \textbf{A recurrence evaluation framework.} We measure how vulnerabilities persist across features, prompt rephrasings, domains, and cross-domain transfer, so that attack success can be linked to model behavior rather than single prompts.
    \item \textbf{An empirical characterization of model-induced risk.} We show that many insecure implementations recur systematically across applications, implying that model-centric security evaluation is necessary in addition to conventional code scanning.
\end{itemize}

Beyond held-out attack success, we find that recurrence survives prompt paraphrase and moderate distribution shift. This means the attack is not merely memorizing a narrow benchmark slice; it exploits stable implementation habits that persist across distinct applications. The broader implication is that evaluating code-model security one sample at a time misses a predictive layer of risk that appears only when generations are analyzed collectively.

\section{Background and Related Work}
\label{sec:related}

\paragraph{Code LLM Evolution and Security Benchmarks}
LLMs have significantly advanced automated code generation, evolving from basic synthesis to complex agentic workflows and AI-native IDEs \cite{chen2021evaluating, li2022competition, nijkamp2022codegen, fried2022incoder, wang2021codet5, guo2022unixcoder, li2023starcoder, roziere2023code}. Despite these functional gains, generated code frequently contains high-impact vulnerabilities in realistic settings \cite{pearce2025asleep}, prompting the creation of specialized security datasets and evaluation benchmarks \cite{siddiq2022securityeval, tony2023llmseceval, bhatt2023purple, wang2024your, peng2025cweval}. Newer end-to-end benchmarks such as WebGenBench and E2EDev move the field closer to realistic application generation, where models construct multi-file systems rather than isolated functions \cite{lu2026webgen, liu2025e2edev}. Even so, most benchmark protocols still summarize security at the level of individual programs: the unit of analysis is whether one sample is vulnerable, not whether the model repeatedly associates the same visible feature with the same hidden flaw.

This per-sample framing limits what current evaluations can reveal. Post-hoc scanners are essential for finding bugs in a specific artifact, but they do not directly measure whether vulnerabilities recur systematically across generations. As a result, they under-characterize a deployment risk that matters in practice: if the same model is used to generate many applications, repeated implementation habits can make hidden vulnerabilities predictable before source-code access is obtained.

\paragraph{Vulnerability Persistence and Fingerprinting}
In the broader machine learning context, black-box security research has demonstrated that probabilistic sampling allows for model extraction and fingerprinting, where stochastic outputs leave stable, identifiable signatures \cite{tramer2016stealing, carlini2021extracting, yang2024fingerprint, jagielski2020high, shokri2017membership}. In LLMs, limited output diversity and structural template reuse have been linked to reduced creativity and persistent ``vulnerability fingerprints'' \cite{elgedawy2024ocassionally, yun2025price}. These results suggest that generation is stable enough not only for model identification, but also for inferring properties of unseen outputs.

Our work applies this intuition to software security. Rather than fingerprinting a model from its raw text, we fingerprint it from the distribution of backend vulnerabilities conditioned on observable frontend features. This is a different level of abstraction: the signal is not a stylistic token pattern, but a recurring mapping between user-visible workflows and hidden security failures. In doing so, we bridge the gap between code-security benchmarking and black-box model fingerprinting.

Classical web-security reconnaissance already uses public workflows to prioritize likely backend issues, but those priors are usually domain heuristics crafted by experts. FSTab instead learns model-specific empirical priors from a profiling corpus. The result is a reusable attack surface that is tied to the generator itself, not only to the application domain or framework.
\section{Feature-Security Table (FSTab)}
\label{sec:method}

FSTab is a model-specific lookup table $\mathcal{T}_m$ that maps an observable frontend feature $f$ to a short ranked list of backend vulnerability rules. The attack assumes a black-box adversary who can interact with a deployed application, identify the model family used to generate it, and observe public UI workflows, routes, and endpoints. The attacker does not require source-code access, provider-side internals, or model weights.

\paragraph{Threat model}
The adversary's capabilities are intentionally modest. More precisely, the attack has two stages. Stage 1 is offline and attacker-controlled: the attacker queries the same publicly accessible model family used by the victim with their own tasks, generates code they own, and builds FSTab once using any source-level analysis they choose. This one-time construction asset is reusable across targets and independent of any particular victim deployment. Stage 2 is online and directed at the victim: the attacker observes only the deployed application's public interface its UI, DOM, and network-visible behavior maps those observations into the 59-action taxonomy, and queries the precomputed table. In this sense, ``black-box'' clarifies the victim-facing inference stage for this class of attack rather than restricting the offline construction stage, and the threat therefore arises from recurring model behavior rather than from compromise of the model provider or privileged access to the deployment stack.

Prior work on LLM fingerprinting suggests that model identity can often be inferred from observable interaction patterns or outputs even in black-box settings \cite{yuan2025llmap, yang2024fingerprint}.

\subsection{Semantic feature extraction}
We define a taxonomy of 59 standardized UI actions that serves as the observable feature vocabulary for both training and attack-time inference. The taxonomy was deliberately designed around user-facing operations login, register, search filtering, photo upload, chat send, payment, order-status lookup, and admin panel access rather than backend-only implementation details. This makes the labels intrinsically UI-observable and portable across the two attack stages: source-level evidence assigns them during offline construction for precision, while UI, DOM, or network-visible evidence supports the same labels during online inference. The goal is to normalize heterogeneous implementations into a small, semantically meaningful action space. Additional taxonomy details and a worked extraction example are provided in Appendix~\ref{app:action} and Appendix~\ref{app:feature_extraction_example}.

For clarity, the source-level extractor used in our pipeline is an offline labeling tool for the attacker-owned construction corpus and for benchmark reproducibility. It should not be read as an online assumption about the victim. At deployment time, the attack consumes only the observable action set; Appendix~\ref{app:attack_demo_fstab} illustrates this black-box identification step from UI interaction alone.

For each scanner finding, we isolate the enclosing function or route context and score candidate actions using lexical and structural evidence from function names, identifiers, string literals, route tokens, and library-specific API patterns. Let $C_l$ denote the local code context around a flagged line $l$. For a candidate action $a$, the extractor computes
\begin{equation}
\label{eq:action_score_main}
\begin{aligned}
S(a, C_l) &= \sum_{k \in \mathcal{K}_a} w_{fn}\mathbb{I}(k \in FN) + w_{id}\mathbb{I}(k \in I) \\
&\qquad + w_{str}\mathbb{I}(k \in L) + w_{route}\mathbb{I}(\mathcal{K}^{r}_a \cap R \neq \emptyset) \\
&\quad + w_{api}\mathbb{I}(\exists r \in \mathcal{R}_a : r \in C_l) \\
&\quad - w_{neg}\lvert \mathcal{N}_a \cap (I \cup L) \rvert,
\end{aligned}
\end{equation}
where $FN$, $I$, $L$, and $R$ are the function-name, identifier, literal, and route-token sets extracted from $C_l$. We use calibrated weights that prioritize semantically reliable signals such as function names and routes over weaker lexical matches, and assign the finding to the highest-scoring action.

\begin{table}[H]
\centering
\small
\caption{Selected frontend features and representative signals from the 59-action taxonomy.}
\label{tab:taxonomy_main}
\begin{tabular}{p{0.18\linewidth}p{0.74\linewidth}}
\toprule
\textbf{Category} & \textbf{Example features and signals} \\
\midrule
Authentication & \texttt{user\_login}\newline keywords: login, authenticate, verify\_credentials\newline routes: \texttt{/auth/login}, \texttt{/signin} \\
Payment & \texttt{submit\_payment}\newline keywords: stripe, checkout, charge, transaction\newline routes: \texttt{/checkout}, \texttt{/pay} \\
Security & \texttt{auth\_token}\newline keywords: jwt, token, bearer, sign, verify\newline libraries: \texttt{jsonwebtoken}, \texttt{pyjwt} \\
Input & \texttt{upload\_file}\newline keywords: upload, attachment, multipart, form\_data\newline exclusion terms separate generic uploads from avatars/images \\
\bottomrule
\end{tabular}
\end{table}

\subsection{Construction and inference}
For each model $m$, we generate a construction corpus of applications, run CodeQL and Semgrep to obtain vulnerability findings, and assign each finding one of the standardized UI actions above. We then aggregate feature-rule co-occurrence counts and rank rules per feature using a PMI-style association score with a diversity penalty so that globally common scanner rules do not dominate every feature.

Let $C(f,r)$ denote the number of times feature $f$ co-occurs with rule $r$ in the construction corpus. We estimate a smoothed conditional probability and combine it with a reuse penalty that discourages generic rules from dominating every feature:
\begin{align}
\hat{P}(r\mid f) &= \frac{C(f,r)+\alpha}{C(f)+\alpha|\mathcal{R}|}, \\
s_{\mathrm{adj}}(f,r) &= \log \frac{\hat{P}(r\mid f)}{\hat{P}(r)} - \log(1+\lambda U[r]),
\end{align}
where $\hat{P}(r)$ is the smoothed marginal frequency of rule $r$ and $U[r]$ counts how often $r$ has already been selected across features. In practice we use additive smoothing $\alpha=0.5$, a diversity penalty $\lambda=0.8$, a maximum list size of $k=25$, and a minimum-support threshold $C(f,r)\geq 3$. On the training split, $k=25$ fully enumerates candidates for 98.6\% of features while still bounding the long-tail cases.

Ground-truth labels come from a dual-engine static-analysis pipeline. We run CodeQL and Semgrep over the generated backends and validate label fidelity through a manual audit of a random subset; the detailed audit protocol and quantitative results are reported in Appendix~\ref{appendix:label_validation}.

Table~\ref{tab:fstab_example_main} shows a representative slice of the learned mapping. Even this small example illustrates the core idea: once the model repeatedly implements a visible feature with the same insecure backend pattern, the feature becomes a useful black-box predictor of hidden risk.

\begin{table}[ht]
\centering
\small
\caption{Illustrative rows from the Claude-4.5 Opus FSTab. The full table appears in Appendix~\ref{table:fstab_claude}.}
\label{tab:fstab_example_main}
\begin{tabular}{p{0.34\linewidth}p{0.56\linewidth}}
\toprule
\textbf{Observed feature} & \textbf{Top recurring vulnerabilities} \\
\midrule
Save New Record To DB & \texttt{py/sql-injection}, \texttt{py/stack-trace-exposure} \\
Upload Document or File & \texttt{py/stack-trace-exposure}, \texttt{js/xss-through-dom} \\
User Login (Password) & \texttt{js/remote-prop-injection}, \texttt{js/insecure-random} \\
Checkout or Payment & \texttt{py/sql-injection}, \texttt{js/missing-rate-limiting} \\
\bottomrule
\end{tabular}
\end{table}

At attack time, the adversary enumerates visible workflows such as login, search, upload, or checkout, maps them to a feature set $F_{\mathrm{obs}}$, and queries the model-specific table:
\begin{equation}
V_{\mathrm{pred}} = \bigcup_{f \in F_{\mathrm{obs}}} \mathcal{T}_m[f].
\end{equation}
The returned rule IDs form a compact set of likely hidden scanner-detected vulnerabilities. FSTab is not intended to certify end-to-end exploitability from black-box evidence alone; it prioritizes which vulnerability classes are most plausible, after which a defender or attacker may validate reachability manually or dynamically. In our experiments, the deduplicated prediction set averages only 4.86 CodeQL rules and 8.03 Semgrep rules per project, making manual validation or downstream automated probing feasible without approximating the full scanner universe. Figure~\ref{fig:e2e_main} summarizes this workflow end to end.

\subsection{Evaluation}
We report three attack metrics. For a target program $P$, let $V_{\mathrm{pred},P}$ be the predicted vulnerability set and $V_{\mathrm{actual},P}$ the scanner-derived ground truth. We define
\begin{equation}
\text{Success}_P=
\begin{cases}
1 & \text{if } |V_{\mathrm{pred},P}\cap V_{\mathrm{actual},P}|>0, \\
0 & \text{otherwise,}
\end{cases}
\end{equation}
and
\begin{align}
\text{Coverage}_P &= \frac{|V_{\mathrm{pred},P}\cap V_{\mathrm{actual},P}|}{|V_{\mathrm{actual},P}|} \\[10pt]
\text{Precision}_P &= \frac{|V_{\mathrm{pred},P}\cap V_{\mathrm{actual},P}|}{|V_{\mathrm{pred},P}|}
\end{align}
Population averages yield ASR, ACR, and APR, where ASR should be read as an overlap-based \emph{scanner-rule hit rate} rather than proof of end-to-end compromise:
\begin{align}
ASR &= \frac{1}{|\mathcal{P}|}\sum_{P\in\mathcal{P}}\text{Success}_P \\[10pt]
ACR &= \frac{1}{|\mathcal{P}|}\sum_{P\in\mathcal{P}}\text{Coverage}_P \\[10pt]
APR &= \frac{1}{|\mathcal{P}|}\sum_{P\in\mathcal{P}}\text{Precision}_P
\end{align}
ASR records whether FSTab surfaces at least one true scanner finding per project. It therefore measures at-least-one overlap between the predicted set and the scanner-derived ground truth, not whether the application is fully compromised. This overlap-based hit criterion is still operationally meaningful for prioritization because a single correct vulnerability class can be enough to guide deeper manual validation, while ACR and APR quantify the breadth and precision of the returned set.

We also measure vulnerability recurrence along four complementary axes: \emph{feature vulnerability recurrence} (FVR), \emph{rephrasing vulnerability persistence} (RVP), \emph{domain vulnerability recurrence} (DVR), and \emph{cross-domain transfer} (CDT). Intuitively, these metrics ask whether the same feature-conditioned vulnerabilities reappear across programs with the same feature, across paraphrases of the same task, within the same application domain, and across different domains. Formal definitions and exhaustive breakdowns are provided in Appendix~\ref{app:formal_metrics}.
\section{Experiments}
\label{sec:experiments}

\subsection{Setup}
We evaluate six SOTA code LLMs: \textbf{Claude~4.5~Opus}, \textbf{Gemini~3~Flash}, \textbf{Gemini~3~Pro}, \textbf{GPT-5.2}, \textbf{Composer}, and \textbf{Grok}. Our primary benchmark is \textbf{WebGenBench} \cite{lu2026webgen}, a larger and more realistic multi-file application benchmark from which we use five domains: E-commerce, Internal Tools, Social Media, Blogging, and Dashboards. We additionally report supplementary transfer results on \textbf{E2EDev} \cite{liu2025e2edev}, a smaller and more template-driven benchmark that is often closer to compact or one-file application patterns. Together these benchmarks yield 1050 generated programs: 900 from WebGenBench and 150 from E2EDev. Within each of these five selected WebGenBench domains: E-commerce, Internal Tools, Social Media, Blogging, and Dashboards. We randomly select 5 prompts for FSTab construction and 5 prompts for held-out testing. Each construction prompt is instantiated as the original instruction plus 4 semantic-preserving rephrasings, while held-out prompts are evaluated only with the original instruction. To test universality, we also evaluate a cross-domain setting in which the target domain is excluded entirely from construction

To measure rephrasing robustness, each task is expanded into five prompt realizations: the original prompt and four semantically equivalent rewrites produced with a few-shot lexical-variation procedure adapted from LM-CPPF \cite{abaskohi2023lm}. This allows us to separate vulnerability persistence from prompt wording. Static-analysis labels come from CodeQL and Semgrep, and label fidelity is discussed in Appendix~\ref{appendix:label_validation}. FSTab predicts likely scanner-detected vulnerabilities rather than certifying individually exploitable bugs, and the attack exploits recurring model behavior across many projects rather than isolated one-off findings.

WebGenBench projects expose a rich visible interface, with 8.6 extracted UI features per project on average. This density makes it possible to ask whether a model's hidden vulnerabilities can be recovered from public functionality alone. Further reproducibility details, model metadata, prompt examples, and full per-domain attack tables are provided in Appendix~\ref{appendix:experiments_section}.

In the evaluation pipeline, we assign these UI-action labels using source-level canonicalization on attacker-owned generations so that large-scale measurements are consistent and reproducible. This should be interpreted as an annotation oracle for benchmark standardization, not as a victim-side requirement: the online attack consumes only the observable action set, which can be recovered manually or with standard UI/DOM/network inspection. As an additional sanity check, we manually verified across the evaluated models and domains that the frontend features assigned by this canonicalization procedure consistently align with the user-visible actions recoverable in this black-box way.

\subsection{Black-box attack results}
\label{sec:attack}

\begin{table*}[ht]
\centering
\small
\caption{Domain-averaged black-box overlap-based prediction performance on WebGenBench. Values are percentages reported as CodeQL $|$ Semgrep. The full per-domain table is deferred to Appendix~\ref{tab:attack_combined_bolded}.}
\label{tab:attack_summary}
\begin{tabular}{lcccc}
\toprule
\textbf{Model} & \textbf{Held-out ASR} & \textbf{Held-out ACR} & \textbf{Cross-domain ASR} & \textbf{Cross-domain ACR} \\
\midrule
GPT-5.2 & 84.00 $|$ 86.00 & 82.00 $|$ 75.83 & 80.92 $|$ 85.42 & 78.62 $|$ 74.66 \\
Claude-4.5 Opus & 85.00 $|$ 88.00 & 85.00 $|$ 87.50 & 78.93 $|$ 85.67 & 78.93 $|$ 84.97 \\
Gemini-3 Pro & 76.00 $|$ 84.00 & 76.00 $|$ 78.13 & 71.31 $|$ 81.00 & 71.31 $|$ 74.13 \\
Gemini-3 Flash & 60.00 $|$ 75.33 & 60.00 $|$ 70.33 & 57.11 $|$ 69.61 & 57.11 $|$ 60.39 \\
Composer & 78.33 $|$ 82.00 & 70.00 $|$ 81.43 & 75.09 $|$ 80.70 & 66.31 $|$ 80.00 \\
Grok & 76.67 $|$ 73.00 & 74.17 $|$ 64.91 & 75.04 $|$ 73.86 & 72.12 $|$ 64.95 \\
\bottomrule
\end{tabular}
\end{table*}

Table~\ref{tab:attack_summary} shows that FSTab frequently recovers at least one held-out scanner finding on unseen targets and remains useful even when the target domain is excluded during construction. GPT-5.2 and Claude-4.5 Opus are strongest on held-out programs, while multiple held-out domains reach perfect ASR and ACR against the scanner-derived labels in the full table reported in the Appendix~\ref{tab:attack_combined_bolded}. Transfer also remains strong: GPT-5.2 retains 80.92$|$85.42 cross-domain ASR, and Claude-4.5 Opus retains 78.93$|$85.67, which shows that the learned vulnerability priors are not merely memorizing one application class.

Precision is predictably lower than ASR because FSTab is optimized to surface at least one plausible scanner-detected weakness per project, not to prove that every returned rule is a confirmed exploit. Gemini-3 Pro attains the highest model-level APR (54.4\%), followed by GPT-5.2 (36.2\%).

\paragraph{Baselines}
To isolate FSTab's contribution from the metric's tolerance to overlap, we compare against a Random-Budget baseline that predicts the same number of rules per project as FSTab, sampled uniformly from the learnable rule space. On WebGenBench held-out, FSTab achieves ASR 69.4\% vs.\ 10.1\% for Random-Budget under CodeQL, and 72.5\% vs.\ 20.3\% under Semgrep, a 3.6$\times$-6.9$\times$ advantage at matched prediction-set size.
Appendix~\ref{app:stats_summary} summarizes the corresponding held-out WebGenBench Wilson intervals and the seed-variation ranges for the stochastic robustness checks.

To test whether FSTab is merely benefiting from a few dominant scanner rules, we recompute ASR and ACR after removing the top-$K$ most frequent training rules from both FSTab's predictions and the scanner-derived ground truth. The matched-budget Random-Budget baseline does not close the gap under this exclusion; if anything, FSTab's relative advantage is preserved or increases. On WebGenBench with Semgrep, removing the top-3 rules leaves FSTab with 63.6\% ASR and 52.7\% ACR on rare-rule projects, versus 18.3\% ASR and 6.9\% ACR for Random-Budget, corresponding to 3.5$\times$ and 7.6$\times$ improvements, respectively. Removing the top-5 rules further widens the long-tail Semgrep gap to 4.0$\times$ in ASR. Under CodeQL, the FSTab/Random-Budget ratio reaches 6.2$\times$ in ASR and 7.6$\times$ in ACR under top-5 exclusion. These results show that FSTab captures predictive structure that extends beyond simple frequency priors and remains informative for less-common vulnerabilities.

FSTab also succeeds with a tightly bounded rule budget. Under the minimum-support filter, it learns only 28 CodeQL rules and 39 Semgrep rules, despite the underlying scanners exposing 300+ and 5{,}000+ rules, respectively. The average query returns only 4.86 CodeQL rules and 8.03 Semgrep rules per project, yet still covers major CWE families including SQL injection, cross-site scripting, path traversal, open redirect, and missing rate limiting.

An additional construction-budget ablation shows that these results do not depend on a large attacker-owned corpus. As detailed in Appendix~\ref{app:budget_ablation}, both ASR and ACR plateau around 20 construction projects, indicating that FSTab already performs strongly under modest budgets.

\subsection{Vulnerability recurrence results}
\label{sec:eval}

\begin{table}[t]
\centering
\small
\caption{Model-level recurrence scores (\%).}
\label{tab:eval_main}
\begin{tabular}{lcccc}
\toprule
\textbf{Model} & \textbf{FVR} $\uparrow$ & \textbf{RVP} $\uparrow$ & \textbf{DVR} $\uparrow$ & \textbf{CDT} $\uparrow$ \\
\midrule
GPT-5.2 & 37.52 & 23.20 & 33.92 & 42.30 \\
Claude-4.5 Opus & 35.37 & 21.44 & 31.75 & 53.58 \\
Gemini-3 Pro & \textbf{51.23} & 25.09 & 41.39 & \textbf{58.67} \\
Gemini-3 Flash & 44.29 & 29.77 & 36.10 & 52.81 \\
Composer & 43.86 & \textbf{35.53} & \textbf{46.43} & 57.32 \\
Grok & 31.43 & 11.96 & 27.85 & 57.29 \\
\bottomrule
\end{tabular}
\end{table}

Table~\ref{tab:eval_main} shows that recurrence is substantial and model dependent. Gemini-3 Pro has the highest feature-level recurrence (FVR=51.23\%), meaning that specific frontend features frequently trigger the same backend flaws. Composer exhibits the strongest persistence under rephrasing (RVP=35.53\%) and the strongest within-domain recurrence (DVR=46.43\%), indicating unusually rigid implementation templates. Cross-domain transfer is strongest for Gemini-3 Pro (CDT=58.67\%), and importantly CDT exceeds DVR for every model. This consistent gap suggests that many vulnerabilities reflect model-level coding habits that generalize across application domains rather than only domain-local artifacts. Full recurrence tables are deferred to Appendix~\ref{appendix:framework_tables}.

Figure~\ref{fig:radar_fingerprints} provides a qualitative view of the same phenomenon. Rather than concentrating in a single task family, recurrence patterns form broad architectural fingerprints that differ by model. This is especially visible for Composer, whose spikes in access control and data operations align with its unusually high RVP and DVR scores.

These fingerprints are not merely descriptive. Each model exhibits its own characteristic distribution of weaknesses, which FSTab encodes as a model-specific vulnerability profile. In a mismatched-fingerprint stress test, we intentionally query each target with the wrong model profile at inference time, sampling uniformly from the other five models and averaging across ten seeds. On WebGenBench under Semgrep, this reduces ACR by 16.4 points on average, corresponding to a 1.36$\times$ degradation relative to the correctly matched setting. The per-model ASR losses can be substantially larger, reaching 25.7 points for Grok, 14.3 for Composer, and 13.8 for Claude-4.5 Opus. These gaps are 4-10$\times$ larger than the per-seed standard deviation, which indicates that the fingerprints are separable, model-specific vulnerability profiles rather than small perturbations of a shared prior.

\begin{table}[ht]
\centering
\small
\caption{Selected slices of the recurrence analysis. Columns report FVR for \textit{Register New Account}, RVP for one representative prompt family, DVR for \textit{E-commerce}, and domain-specific CDT for \textit{E-commerce}.}
\label{tab:eval_detail_main}
\begin{tabular}{lcccc}
\toprule
\textbf{Model} & \textbf{FVR} $\uparrow$ & \textbf{RVP} $\uparrow$ & \textbf{DVR} $\uparrow$ & \textbf{$CDT_d$} $\uparrow$ \\
\midrule
GPT-5.2 & \textbf{100.00} & 49.70 & 46.67 & 73.33 \\
Claude-4.5 Opus & 66.67 & 32.95 & 48.78 & 53.66 \\
Gemini-3 Pro & \textbf{100.00} & 31.16 & 33.33 & 60.00 \\
Gemini-3 Flash & \textbf{100.00} & 31.67 & 37.50 & \textbf{87.50} \\
Composer & 40.00 & \textbf{49.98} & \textbf{50.94} & 56.60 \\
Grok & \textbf{100.00} & 23.68 & 37.04 & 55.56 \\
\bottomrule
\end{tabular}
\end{table}

The selected slices in Table~\ref{tab:eval_detail_main} make the aggregate scores more concrete. Some frontend actions are almost deterministic vulnerability triggers for specific models. Composer's rephrasing persistence approaches 50\%, which indicates that many of its vulnerabilities survive substantial changes in wording. At the same time, Gemini-3 Flash reaches the strongest E-commerce transfer score, showing that domain-excluded construction can still surface useful priors for a target application.

\begin{figure}[H]
    \centering
    \includegraphics[width=\linewidth]{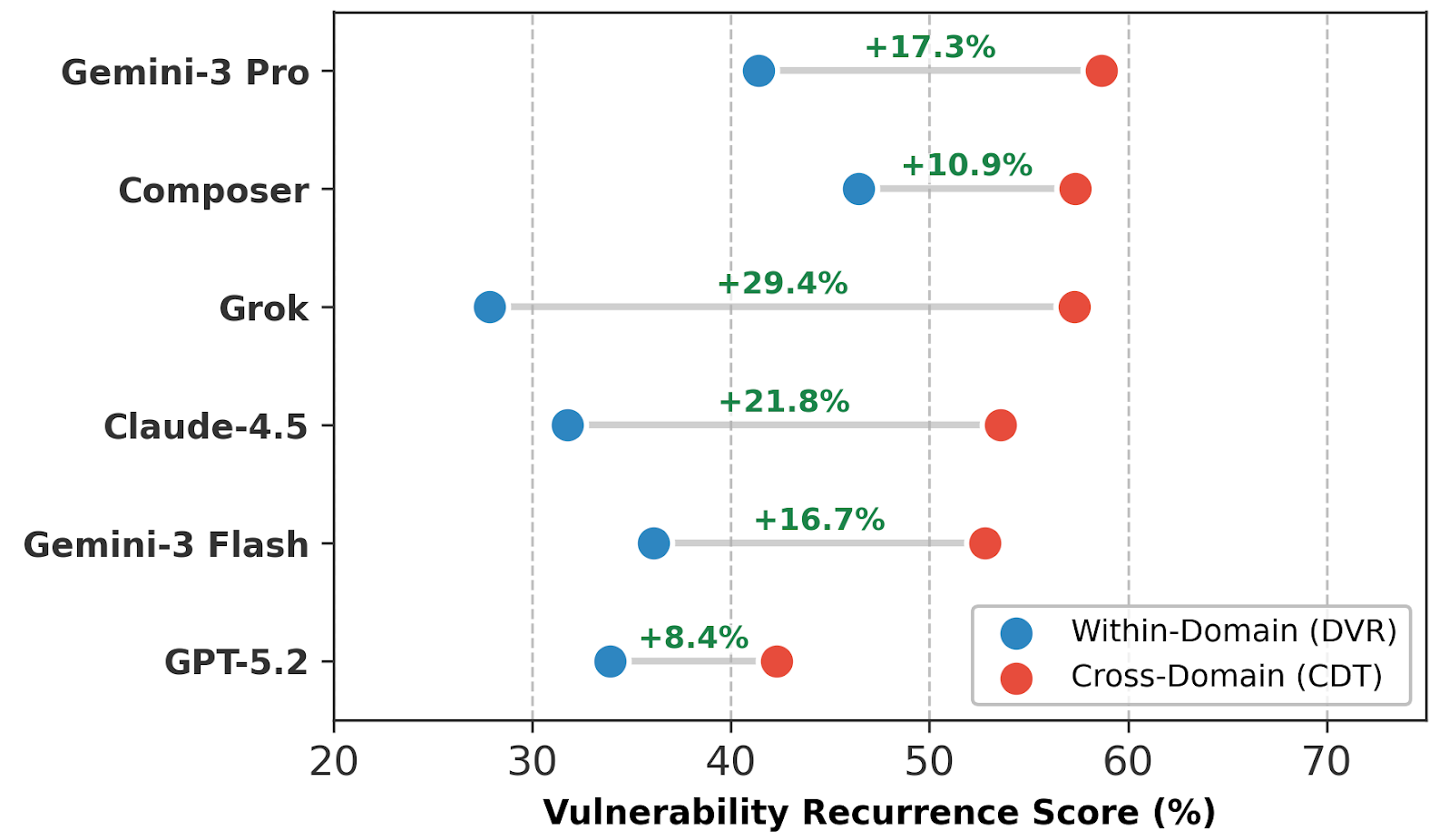}
    \caption{The universality gap. For every model, cross-domain transfer exceeds within-domain recurrence, indicating that many recurring vulnerabilities are model-level coding habits rather than domain-specific artifacts.}
    \label{fig:universality_gap_main}
\end{figure}

The universality gap in Figure~\ref{fig:universality_gap_main} is one of the clearest results in the paper. On average, $CDT$ exceeds $DVR$ by roughly 18 points, which means that vulnerabilities learned from unrelated application domains often transfer more reliably than vulnerabilities constrained to the same domain. This behavior is hard to explain with narrow prompt memorization; it is much more consistent with persistent model-level implementation biases.
\section{Discussion and Limitations}
\label{sec:discussion}

Our results show that visible application features can act as reliable priors over hidden backend vulnerabilities in LLM-generated software. This turns vulnerability recurrence from an evaluation curiosity into a practical black-box attack surface: once a model repeatedly implements the same feature with the same insecure template, an attacker can exploit that regularity without inspecting source code.

The findings also suggest a concrete defensive agenda. If recurring vulnerabilities are tied to specific frontend features, then model providers and downstream integrators can build feature-conditioned regression suites, targeted rewriting passes for high-risk workflows, and security filters that activate when the model generates sensitive functionality such as authentication, file upload, or payment logic. More broadly, model evaluation should track recurring failure modes across generations, not only aggregate vulnerability counts on isolated samples.

FSTab also has clear limitations. It assumes either model identity or a reliable fingerprinting step, and it depends on identifying a target's observable UI actions. In our experiments, these actions are canonicalized with automated extraction on attacker-owned generations for consistency, but this should be understood as an evaluation convenience rather than a victim-side requirement: at deployment time, the same information can be obtained manually or with standard UI/DOM/network inspection. Our benchmarks focus on LLM-generated web applications, so the observable feature space is richer than in smaller script-generation tasks. In addition, FSTab predicts likely scanner-detected vulnerabilities, not guaranteed end-to-end exploits. Static analysis is only a proxy for exploitability, though our manual audit in Appendix~\ref{appendix:label_validation} indicates that most retained findings correspond to genuine insecure patterns (82\% precision for CodeQL and 74\% for Semgrep), and the representative case study in Appendix~\ref{app:attack_demo_fstab} shows that FSTab-guided prioritization can still surface practically risky attack paths.

Even with these constraints, the cross-feature, cross-prompt, and cross-domain regularities are strong enough to motivate model-centric defenses such as security-aware rewriting, domain-held-out evaluation, and targeted red teaming around recurring feature templates. Future work could combine FSTab-style priors with dynamic testing, extend the method to agent-generated repositories and mobile applications, and study whether explicit diversity objectives reduce the predictability of hidden vulnerabilities.

\bibliography{ref}
\bibliographystyle{icml2026}

%%%%%%%%%%%%%%%%%%%%%%%%%%%%%%%%%%%%%%%%%%%%%%%%%%%%%%%%%%%%%%%%%%%%%%%%%%%%%%%
%%%%%%%%%%%%%%%%%%%%%%%%%%%%%%%%%%%%%%%%%%%%%%%%%%%%%%%%%%%%%%%%%%%%%%%%%%%%%%%
% APPENDIX
%%%%%%%%%%%%%%%%%%%%%%%%%%%%%%%%%%%%%%%%%%%%%%%%%%%%%%%%%%%%%%%%%%%%%%%%%%%%%%%
%%%%%%%%%%%%%%%%%%%%%%%%%%%%%%%%%%%%%%%%%%%%%%%%%%%%%%%%%%%%%%%%%%%%%%%%%%%%%%%
\newpage
\appendix
\onecolumn
%%%%%%%%%%%%%%%%%%%%%%%%%%%%%%%%%%%%%%%%%%%%%%%%%%%%%%%%%%%%%%%%%%%%%%%%%%%%%%%
% APPENDIX: Extracting Recurring Vulnerabilities from Black-Box LLM-Generated Software
%%%%%%%%%%%%%%%%%%%%%%%%%%%%%%%%%%%%%%%%%%%%%%%%%%%%%%%%%%%%%%%%%%%%%%%%%%%%%%%

\clearpage

\section*{Ethical Considerations}

Our work studies how recurring vulnerabilities in code-generative LLMs can be inferred and exploited from black-box observations. The primary benefit of this research is defensive: by showing that insecure implementation patterns persist across rephrasings, domains, and model families, we provide evidence that can help benchmark designers, model providers, static-analysis tool builders, and downstream developers identify and mitigate systematic failure modes in LLM-generated software.

The main risk is that our methodology could lower the effort required to identify likely vulnerabilities in applications produced by code-generative models. We therefore structure the study to minimize harm. First, our evaluation is conducted on benchmark tasks and controlled model outputs rather than on live third-party systems, and we do not target real users or collect private user data. Second, we report aggregate recurrence patterns, rule families, and model-level behavior rather than publishing a turnkey exploitation workflow against specific deployed targets. Third, when discussing concrete vulnerabilities, we focus on standard vulnerability classes already captured by widely used scanners (e.g., CodeQL and Semgrep), so the paper emphasizes measurement and diagnosis rather than novel weaponization.

This work does not involve human subjects, personally identifiable information, or user studies, so IRB/ERB review was not required. More broadly, our ethical analysis does not rely solely on institutional compliance: we weigh the risk of misuse against the security value of documenting a predictable attack surface that already exists in production-facing coding assistants. We believe the benefits of exposing these recurring vulnerabilities outweigh the risks, especially because the results motivate safer model deployment, stronger secure-code post-processing, and more realistic evaluation standards for LLM-generated software.

\appendix
\onecolumn % Recommended for complex tables and code snippets

\begin{center}
    {\huge \textbf{Appendix}}
\end{center}

\section{Cost Analysis}

To facilitate reproducibility and provide a realistic resource estimation for the FSTab framework, we detail the financial and computational costs associated with our experiments. Our experimental infrastructure utilized the \textbf{Cursor Ultra} subscription tier to support the high-context requirements of full-stack vulnerability analysis.

\subsection{Infrastructure \& Financial Layout}
The experiments were conducted using a "Long Context" agentic workflow, requiring the entire project codebase to be loaded into the model's context window for accurate feature extraction.
\begin{itemize}
    \item \textbf{Total Financial Cost:} \textbf{\$419.40}
    \item \textbf{Breakdown:}
    \begin{itemize}
        \item \textbf{\$200.00:} Cursor Ultra monthly plan.
        \item \textbf{\$219.40:} Usage-based overage for high-volume inference.
    \end{itemize}
    \item \textbf{Token Volume:} The study generated and analyzed 900 distinct applications. Due to the requirement of analyzing full directory structures, the average context size exceeded \textbf{100,000 tokens per project}.
\end{itemize}

\subsection{Cost Breakdown by Phase}
We categorize the costs into two phases. The "Construction" phase dominated the resource consumption due to the diversity requirements ($K=5$ rephrasings), while the "Test" phase proved highly efficient.

\begin{table}[H]
\centering
\caption{Resource Consumption \& Cost Allocation (Total Budget: \$419.40)}
\label{tab:cost_split}
\begin{tabular}{l c c c | r}
\toprule
\textbf{Model Family} & \textbf{Avg. Tokens} & \textbf{Construction} & \textbf{Test} & \textbf{Allocated Cost} \\
 & \small{(Per Project)} & \small{(125 Apps)} & \small{(50 Apps)} & \small{(USD)} \\
\midrule
\textbf{Claude 4.5 Opus} & $\sim$115k & \$148.00 & \$74.00 & \textbf{\$222.00} \\
\textbf{GPT-5.2} & $\sim$110k & \$48.20 & \$24.10 & \textbf{\$72.30} \\
\textbf{Gemini 3 Pro} & $\sim$125k & \$36.40 & \$18.20 & \textbf{\$54.60} \\
\textbf{Composer 1} & $\sim$105k & \$28.40 & \$14.20 & \textbf{\$42.60} \\
\textbf{Gemini 3 Flash} & $\sim$130k & \$13.20 & \$6.60 & \textbf{\$19.80} \\
\textbf{Grok Code} & $\sim$120k & \$5.40 & \$2.70 & \textbf{\$8.10} \\
\midrule
\textbf{Totals} & \textbf{$>$115k avg} & \textbf{\$279.60} & \textbf{\$139.80} & \textbf{\$419.40} \\
\bottomrule
\end{tabular}
\end{table}

\newpage

\section{Introduction}
\subsection{Overview of the FSTab Workflow}

Figure~\ref{fig:appendix_e2e_overview} provides the end-to-end pipeline that was removed from the main paper for space. It illustrates the construction and attack phases of FSTab: generate applications, label them with security findings and frontend features, build a model-specific lookup table, then query that table from the visible surface of a target application.

\begin{figure}[H]
    \centering
    \includegraphics[width=\linewidth]{imgs/End_to_End.pdf}
    \caption{Overview of the FSTab framework.}
    \label{fig:appendix_e2e_overview}
\end{figure}

\subsection{Figure 1 - Architectural Vulnerability Fingerprints}
\label{app:radar_details}

To translate the abstract statistical metrics of Feature Vulnerability Recurrence (FVR) into interpretable "Security Personalities," we developed the Architectural Vulnerability Fingerprint visualization. This subsection details the data aggregation and construction process used to generate these radar plots.

\begin{figure}[H]
    \centering
    \includegraphics[width=0.5\linewidth]{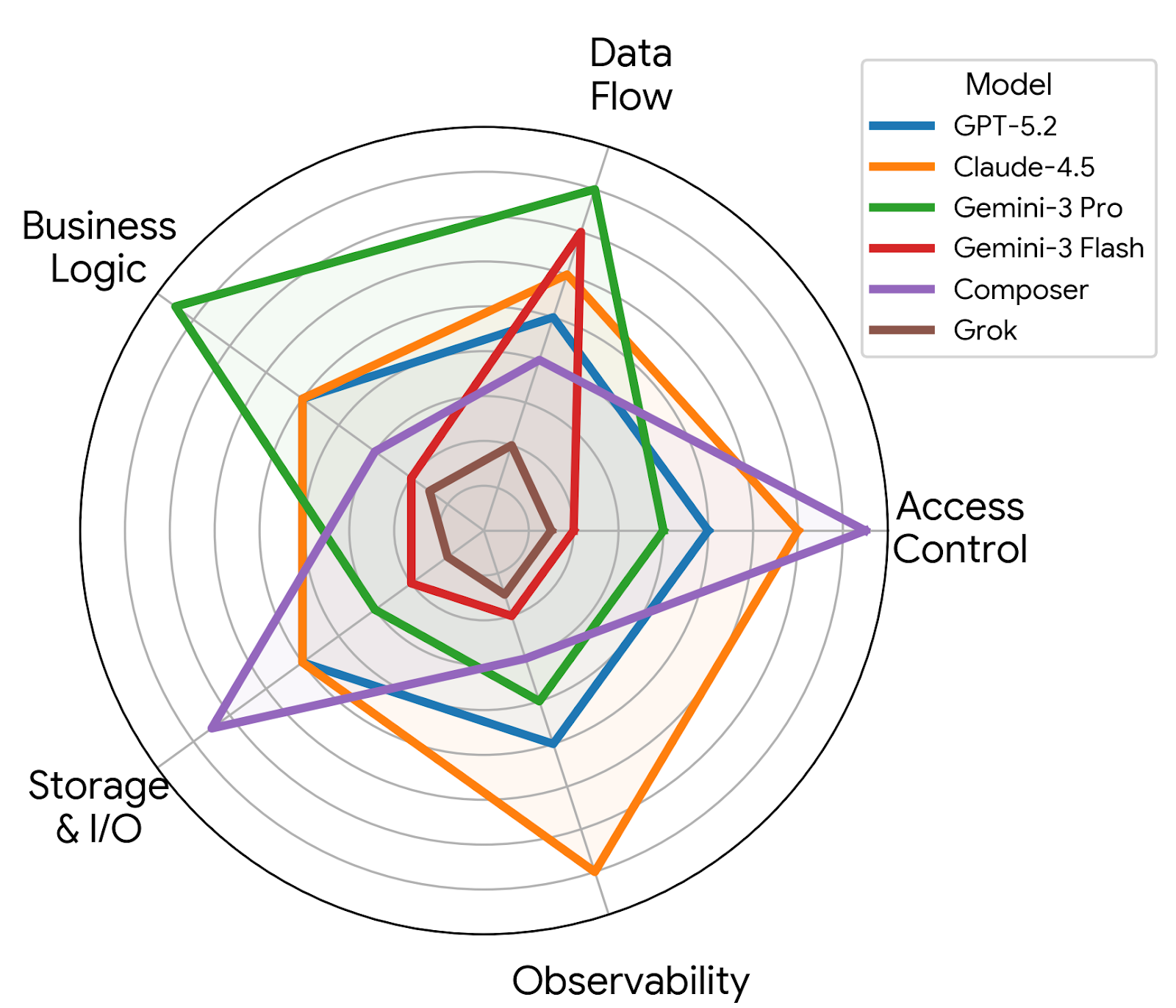}
    \caption{Architectural vulnerability fingerprints.}
    \label{fig:radar_fingerprints_app}
\end{figure}

\paragraph{Methodology \& Taxonomy}
Our raw experimental data consists of FVR scores for $N=59$ distinct frontend features (e.g., \textit{User Login}, \textit{Export CSV}). To visualize high-level architectural biases, we mapped these granular features into five semantic categories representing core software architectural components:
\begin{itemize}
    \item \textbf{Access Control:} Features governing authentication and authorization (e.g., \textit{Register Account}, \textit{Reset Password}). High persistence here indicates the model relies on rigid, potentially insecure templates for security-critical logic.
    \item \textbf{Data Flow:} Features involving data movement and transformation (e.g., \textit{File Upload}, \textit{Download Report}). Recurrence suggests persistence in I/O handling.
    \item \textbf{Business Logic:} Features executing core domain rules (e.g., \textit{Calculate Tax}, \textit{Process Transaction}). Spikes here reveal rigid algorithmic implementations.
    \item \textbf{Storage \& I/O:} Features interacting directly with persistence layers (e.g., \textit{Save Record}, \textit{Delete Entry}).
    \item \textbf{Observability:} Features related to monitoring (e.g., \textit{View Analytics}, \textit{System Logs}).
\end{itemize}

\paragraph{Score Computation}
For each model $M$ and category $C$, the architectural recurrence score $S_{M,C}$ is calculated as the mean FVR of all features $f$ belonging to that category: $S_{M,C} = \frac{1}{|F_C|} \sum_{f \in F_C} FVR(M, f)$. The resulting scores are normalized on a radial axis from $0.0$ (Stochastic) to $1.0$ (Systematic Recurrence).

\paragraph{Inference}
The resulting geometric shapes allow for rapid visual inference of a model's security posture. A sharp spike along a specific axis (e.g., \textbf{Composer}'s protrusion in \textit{Access Control}) indicates that the model is "over-fitted" to a specific coding pattern for that domain. The total area enclosed by the fingerprint correlates with the model's overall susceptibility to FSTab—larger areas (e.g., \textbf{Gemini-3 Pro}) reflect broad rigidity, while minimal areas (e.g., \textbf{Grok}) indicate high entropy and stochasticity.

\section{Method Details}
\label{appendix:method_details}

\subsection{Semantic Action Extraction and Taxonomy}
\label{semantic_action_extraction}

We define a taxonomy of 59 UI actions ($\mathcal{A}$) serving as observable frontend features. While labeled via structural parsing during development for precision, these categories represent functionality natively visible through public UI and API endpoints during black-box inference.

We extract these features from the source code $\mathcal{C}$ at specific line numbers $l$ using a heuristic scoring function that combines lexical analysis with structural parsing.

\paragraph{Structural Context Extraction}
We first isolate the relevant code context, $C_l$, surrounding line $l$.
\begin{itemize}
    \item \textbf{Python:} We utilize Abstract Syntax Tree (AST) parsing to identify the innermost function or class enclosure, explicitly extracting decorator arguments (e.g., \texttt{@app.route('/login')}) to capture routing metadata.
    \item \textbf{JavaScript/TypeScript:} We employ a combination of Tree-sitter parsing and regular expressions to identify function boundaries and route definitions (e.g., \texttt{app.post('/api/pay', ...)}).
\end{itemize}

\paragraph{Action Scoring}
For a candidate action $a \in \mathcal{A}$ and context $C_l$, we calculate a relevance score $S(a, C_l)$. The score aggregates evidence from identifiers ($I$), string literals ($L$), function names ($FN$), route tokens ($R$), and API usage patterns ($R_{api}$):
\begin{equation}
\label{app:score_func}
\begin{aligned}
S(a, C_l) &= \sum_{k \in \mathcal{K}_a} \left( w_{fn} \mathbb{I}(k \in FN) + w_{id} \mathbb{I}(k \in I) + w_{str} \mathbb{I}(k \in L) \right) \\
&\quad + w_{route} \mathbb{I}(\mathcal{K}^r_a \cap R \neq \emptyset) + w_{api} \mathbb{I}(\exists r \in \mathcal{R}_a : r \in C_l) \\
&\quad - w_{neg} \cdot |\mathcal{N}_a \cap (I \cup L)|.
\end{aligned}
\end{equation}
Here $\mathcal{K}_a$ denotes the keyword set for action $a$, $\mathcal{K}^r_a$ its route keywords, $\mathcal{R}_a$ its API regex patterns, and $\mathcal{N}_a$ its negative keywords. The identifier and string contributions are capped at $\tau_{id}=1.5$ and $\tau_{str}=1.0$.

\paragraph{Weight Selection}
The calibrated weights reflect the semantic reliability of each evidence source: $w_{fn}=2.5$ for function names, $w_{route}=2.0$ for route definitions, $w_{api}=1.8$ for library-specific API patterns, $w_{id}=0.5$ for general identifiers, $w_{str}=0.35$ for string literals, and $w_{neg}=0.5$ for negative keywords. The final assignment is $\hat{a}=\arg\max_a S(a,C_l)$.

\subsubsection{Action Taxonomy}
\label{app:action}

Table~\ref{tab:taxonomy_examples} illustrates a subset of the semantic action inventory.

\begin{table}[H]
\centering
\small
\caption{Selected frontend features and keyword signals.}
\label{tab:taxonomy_examples}
\begin{tabular}{p{0.18\linewidth}p{0.74\linewidth}}
\toprule
\textbf{Category} & \textbf{Example features \& signals} \\
\midrule
\textbf{Authentication} & \texttt{user\_login\_with\_password} \newline \textit{Keywords:} login, authenticate, verify\_credentials \newline \textit{Routes:} \texttt{/auth/login}, \texttt{/signin} \\
\textbf{Payment} & \texttt{submit\_payment} \newline \textit{Keywords:} stripe, checkout, charge, transaction \newline \textit{Routes:} \texttt{/checkout}, \texttt{/pay} \\
\textbf{Data Access} & \texttt{fetch\_data\_from\_database} \newline \textit{Keywords:} get, fetch, retrieve, view \newline \textit{HTTP Methods:} GET (inferred via heuristic) \\
\textbf{Security} & \texttt{generate\_or\_validate\_auth\_token} \newline \textit{Keywords:} jwt, token, bearer, sign, verify \newline \textit{Libs:} \texttt{jsonwebtoken}, \texttt{pyjwt} \\
\textbf{Input} & \texttt{upload\_document\_or\_file} \newline \textit{Keywords:} upload, attachment, multipart, form\_data \newline \textit{Exclusions:} image, avatar (mapped to distinct features) \\
\bottomrule
\end{tabular}
\end{table}

\paragraph{Vulnerability Scanner}
To generate ground-truth labels, we use CodeQL and Semgrep over the backend source code of the construction corpus. The output constitutes the set of actual vulnerabilities used to train FSTab. Manual verification of a random subset confirms that the automated labels accurately reflect present code vulnerabilities (Appendix~\ref{appendix:label_validation}).

\subsection{Hyperparameter Sensitivity for FSTab Construction}
\label{appendix:fstab_hparams}

FSTab has three core hyperparameters: smoothing $\alpha$, list size $k$, and diversity penalty $\lambda$. These act as regularizers controlling a bias--variance trade-off: stabilizing PMI under sparse counts, bounding the candidate budget per feature, and preventing concentration on globally frequent rules.

\textbf{Smoothing ($\alpha=0.5$).} PMI is high-variance when $C(f,r)$ is small because it contains a log-ratio of empirical probabilities. We therefore use a conservative half-count prior to stabilize low-count pairs without flattening the distribution too aggressively.

\textbf{Top-$k$ list size ($k=25$).} A useful proxy for candidate coverage is
\begin{equation}
\label{eq:covk_app}
\mathrm{Cov}_k(f) =
\frac{\sum_{r\in \mathcal{T}[f]} \hat{P}(r\mid f)}{\sum_{r:C(f,r)>0} \hat{P}(r\mid f)}.
\end{equation}
On the training split, most features co-occur with only a small number of distinct rules: the number of candidate rules per feature has p25/median/p75 of $2/3/5$ and a maximum of $41$. Thus $k=25$ fully enumerates candidates for $98.6\%$ of features while still bounding the long-tail cases.

\textbf{Diversity penalty ($\lambda=0.8$).} The diversity term discourages a small set of generic scanner rules from dominating every feature. After a rule has been selected $m$ times, its effective multiplicative discount is $\gamma(m;\lambda)=1/(1+\lambda m)$. With $\lambda=0.8$, the first reuse reduces a rule's effective ratio to $0.56$, and after five uses to $0.20$, which is strong enough to suppress super-nodes without eliminating genuinely cross-cutting vulnerabilities.

\begin{table}[H]
\caption{Training-split sensitivity for $k$ and $\lambda$. Concentration uses the top-1 rule per feature (MaxUse$_1$, Gini$_1$).}
\label{tab:fstab_hparam}
\centering
\resizebox{\textwidth}{!}{%
\begin{tabular}{@{}c c c|c c c c@{}}
\toprule
\multicolumn{3}{c|}{$\lambda=0.8$} & \multicolumn{4}{c}{$k=25$} \\
\cmidrule(lr){1-3}\cmidrule(lr){4-7}
$k$ & Mean $\mathrm{Cov}_k \uparrow$ & Avg $|\mathcal{M}[f]|$ & $\lambda$ & Mean $\mathrm{Cov}_{25} \uparrow$ & MaxUse$_1 \downarrow$ & Gini$_1 \downarrow$ \\
\midrule
5  & 0.866 & 3.145 & 0.0 & 0.988 & 10 & 0.559 \\
10 & 0.960 & 4.000 & 0.2 & 0.987 & 8 & 0.567 \\
20 & 0.986 & 4.362 & 0.4 & 0.987 & 8 & 0.555 \\
25 & 0.987 & 4.435 & 0.8 & 0.987 & 8 & 0.555 \\
\bottomrule
\end{tabular}%
}
\end{table}

\subsection{Formal Attack and Recurrence Metrics}
\label{app:formal_metrics}

\paragraph{Attack Evaluation Metrics}\label{sec:attack_definitions}
For a program $P$, let $V_{\text{pred},P}$ denote the vulnerabilities predicted by FSTab and $V_{\text{actual},P}$ the scanner-derived ground truth. We define
\begin{equation}
\text{Success}_P =
\begin{cases}
1 & \text{if } |V_{\text{pred},P} \cap V_{\text{actual},P}| > 0, \\
0 & \text{otherwise,}
\end{cases}
\end{equation}
\begin{equation}
\text{Coverage}_P = \frac{|V_{\text{pred},P} \cap V_{\text{actual},P}|}{|V_{\text{actual},P}|},
\end{equation}
and
\begin{equation}
\text{Precision}_P = \frac{|V_{\text{pred},P} \cap V_{\text{actual},P}|}{|V_{\text{pred},P}|}.
\end{equation}
Population-level averages give ASR, ACR, and APR, where ASR should be interpreted as an at-least-one static-finding overlap metric:
\begin{equation}
ASR = \frac{1}{|\mathcal{P}|} \sum_{P \in \mathcal{P}} \text{Success}_{P},
\end{equation}
\begin{equation}
ACR = \frac{1}{|\mathcal{P}|} \sum_{P \in \mathcal{P}} \text{Coverage}_{P},
\end{equation}
\begin{equation}
APR = \frac{1}{|\mathcal{P}|} \sum_{P \in \mathcal{P}} \text{Precision}_{P}.
\end{equation}

\paragraph{Recurrence Framework}\label{sec:metrics}
Let $M$ be an LLM and let $P$ denote the set of programs generated by $M$ under our benchmark. Programs are indexed by $(p,k)$, where $p \in \mathcal{P}$ denotes a code-generation task and $k \in \{1,\dots,K\}$ denotes a semantic-preserving rephrasing. Each program $P_{p,k}$ has a domain label $d(p) \in D$, a set of frontend features $F_{p,k} \subseteq \mathcal{F}$, and a set of backend security findings. A vulnerability is defined as a feature--security-rule pair $(f,v)$ and is counted at most once per program.

Let $\mathcal{V}_{p,k}$ denote the set of distinct $(f,v)$ vulnerabilities observed in program $P_{p,k}$. Given a group of programs $G \subseteq P$, define
\begin{equation}
\mathrm{freq}_G(f,v) = \left| \{ P_{p,k} \in G \mid (f,v) \in \mathcal{V}_{p,k} \} \right|,
\end{equation}
and
\begin{equation}
\mathrm{Rec}(G) =
\frac{\left| \{(f,v) \mid \mathrm{freq}_G(f,v) > 1 \} \right|}{\left| \{(f,v) \mid \mathrm{freq}_G(f,v) \ge 1 \} \right|}.
\label{eq:rec}
\end{equation}

\paragraph{(1) Feature Vulnerability Recurrence (FVR)}\label{sec:fvr}
For each feature $f$, let $G_f$ be the set of programs containing $f$. The model-level FVR is
\begin{equation}
FVR_{\text{model}} = \frac{1}{|\mathcal{F}|} \sum_{f \in \mathcal{F}}
\frac{\left| \{(f,v) \mid \mathrm{freq}_{G_f}(f,v) > 1 \} \right|}{\left| \{(f,v) \mid \mathrm{freq}_{G_f}(f,v) \ge 1 \} \right|}.
\end{equation}

\paragraph{(2) Rephrasing Vulnerability Persistence (RVP)}\label{sec:rvp}
For a task $p$, let $\mathcal{V}_{p}^{(k)} = \mathcal{V}_{p,k}$, define $\mathcal{V}_{p}^{\cup} = \bigcup_{k=1}^{K} \mathcal{V}_{p}^{(k)}$, and let
\begin{equation}
\mathcal{V}_{p}^{\cap} = \{(f,v) \mid |\{k:(f,v) \in \mathcal{V}_{p}^{(k)}\}| > 1\}.
\end{equation}
Then
\begin{equation}
RVP_p = \frac{|\mathcal{V}_{p}^{\cap}|}{|\mathcal{V}_{p}^{\cup}|},
\end{equation}
and
\begin{equation}
RVP_{\text{model}} = \frac{1}{|\mathcal{P}|} \sum_{p\in\mathcal{P}} RVP_p.
\end{equation}

\paragraph{(3) Domain Vulnerability Recurrence (DVR)}\label{sec:dvr}
For a domain $d$, let $G_d = \{P_{p,k} \in P \mid d(p)=d\}$. Then
\begin{equation}
DVR_d = \frac{\left| \{(f,v) \mid \mathrm{freq}_{G_d}(f,v) > 1 \} \right|}{\left| \{(f,v) \mid \mathrm{freq}_{G_d}(f,v) \ge 1 \} \right|},
\end{equation}
and
\begin{equation}
DVR_{\text{model}} = \frac{1}{|D|} \sum_{d \in D} DVR_d.
\end{equation}

\paragraph{(4) Cross-Domain Vulnerability Transfer (CDT)}\label{sec:cdt}
For a target domain $d$, let
\begin{equation}
\mathcal{V}_d = \{(f,v) \mid \exists (p,k)\ \text{s.t.}\ d(p)=d,\ (f,v)\in\mathcal{V}_{p,k}\},
\end{equation}
and
\begin{equation}
\mathcal{V}_{\neg d} = \{(f,v) \mid \exists (p,k)\ \text{s.t.}\ d(p)\neq d,\ (f,v)\in\mathcal{V}_{p,k}\}.
\end{equation}
Then
\begin{equation}
CDT_d = \frac{|\mathcal{V}_d \cap \mathcal{V}_{\neg d}|}{|\mathcal{V}_d|},
\end{equation}
and
\begin{equation}
CDT_{\text{model}} = \frac{1}{|D|}\sum_{d\in D} CDT_d.
\end{equation}

\section{Experiments}
\label{appendix:experiments_section}
\subsection{Reproducibility Checklist \& Infrastructure}
\label{appendix:reproducibility}
To ensure the reproducibility of our results (Section~\ref{sec:experiments}), we provide the following details:
\begin{itemize}
    \item \textbf{Compute Resources:} All experiments and model inferences were conducted locally on a workstation equipped with an Apple M2 Max processor.
    \item \textbf{Software Environment:} The agentic workflow was implemented and executed using the Cursor IDE, leveraging its integrated AI agent capabilities for code generation and modification.
    \item \textbf{Vulnerability Scanners:} Ground-truth labels ($V_{actual,P}$) were obtained using CodeQL version 2.23.8  and Semgrep version 1.147.0 with standard security rule sets.
\end{itemize}

\subsection{Experimental Settings}
\subsubsection{Models}
\label{appendix:model_metadata}
We evaluated six models. Below are the specific versions used for the evaluation corpora generation.

\begin{table}[H]
\centering
\caption{LLM API Versions and Documented Providers.}
\label{table:model_metadata}
\begin{tabular}{lll}
\toprule
\textbf{Model Family} & \textbf{Specific Version} & \textbf{Provider} \\ \midrule
GPT-5 & \texttt{gpt-5.2} & OpenAI\\
Claude & \texttt{claude-4.5-opus} & Anthropic\\
Gemini & \texttt{gemini-3-pro} & Google\\
Gemini & \texttt{gemini-3-flash} & Google\\
Composer & \texttt{composer-1} & Cursor\\
Grok & \texttt{grok-code} & xAI \\ 
\bottomrule
\end{tabular}
\end{table}

\subsubsection{Prompt Rephrasing Methodology}
\label{appendix:prompt_samples}

To evaluate RVP, we construct for each task a set of five prompt realizations: the original prompt ($P_0$) and four semantically equivalent but syntactically distinct rephrasings ($P_1$--$P_4$). We use a few-shot in-context procedure adapted from LM-CPPF \citep{abaskohi2023lm} to encourage lexical and syntactic variation while preserving task semantics and avoiding hallucinated or omitted requirements.

\paragraph{Representative example}
Below is a representative example from the ``Big Data Visualization Platform'' task. Although the wording changes substantially across $P_1$--$P_4$, the functional requirements remain invariant.

\begin{itemize}
    \item \textbf{Base Prompt ($P_0$):} ``Please implement a big data visualization platform for data analysis and display. The platform should have functionalities for data visualization, report editing, dashboard preview, resource tree management, text editing, and data analysis. Users should be able to log in, browse data, edit reports, preview dashboards, manage resource trees, edit text, and perform data analysis. Use light salmon for the background and coral for components.''
    \item \textbf{Rephrasing 1 ($P_1$):} ``Create a platform for big data visualization and analysis. It must include features for editing reports, viewing dashboards, managing resource trees, and editing text. Users should be able to log in, explore data, and analyze it. Set the background to light salmon and components to coral.''
    \item \textbf{Rephrasing 2 ($P_2$):} ``Develop a big data analysis and display system. Key functions should be data visualization, report creation, dashboard previews, and resource management. Users need to log in to browse data and perform analyses. Use light salmon for the background color and coral for the elements.''
    \item \textbf{Rephrasing 3 ($P_3$):} ``I need a solution for visualizing and analyzing big data. The system should allow users to log in, manage resources, edit text and reports, and preview dashboards. The design must use a light salmon background with coral-colored components.''
    \item \textbf{Rephrasing 4 ($P_4$):} ``Construct a web platform dedicated to big data visualization. Features include report editing, data analysis, and dashboard previews. Users must be able to log in and manage data. The UI should feature a light salmon background and coral components.''
\end{itemize}

This separation of semantic invariants from surface wording allows us to test whether insecure generations are robust failure modes or artifacts of a particular phrasing.

\subsection{FSTab Attack Evaluation}
\subsubsection{Full WebGenBench Attack Results}
\label{tab:attack_combined_bolded}

Table~\ref{tab:attack_combined_bolded} reports the full per-domain attack table that was removed from the main paper in favor of a compact domain-averaged summary.

\begin{table}[H]
\centering
\caption{Overlap-based prediction performance on the WebGenBench dataset. We report CodeQL and Semgrep results for ASR and ACR (\%). ASR denotes at-least-one overlap with the scanner findings. Values are presented as CodeQL | Semgrep.}

\resizebox{\textwidth}{!}{
\begin{tabular}{lcccccccccc}
\toprule
\textbf{Model} &
\multicolumn{2}{c}{\textbf{E-commerce}} &
\multicolumn{2}{c}{\textbf{Internal Tools}} &
\multicolumn{2}{c}{\textbf{Social Media}} &
\multicolumn{2}{c}{\textbf{Blogging}} &
\multicolumn{2}{c}{\textbf{Dashboards}} \\
& ASR $\uparrow$ & ACR $\uparrow$ & ASR $\uparrow$ & ACR $\uparrow$ & ASR $\uparrow$ & ACR $\uparrow$ & ASR $\uparrow$ & ACR $\uparrow$ & ASR $\uparrow$ & ACR $\uparrow$ \\
\midrule
\multicolumn{11}{l}{\textit{\textbf{Held-out (target-domain) evaluation}}} \\
\midrule
GPT-5.2 & \textbf{100} | \textbf{100} & \textbf{100} | 86.67 & \textbf{60} | \textbf{75} & \textbf{60} | \textbf{75} & \textbf{80} | 75 & 70 | 52.50 & 80 | 80 & 80 | 65 & \textbf{100} | \textbf{100} & \textbf{100} | \textbf{100} \\
Claude-4.5 Opus & \textbf{100} | \textbf{100} & \textbf{100} | \textbf{100} & 50 | 60 & 50 | 60 & 75 | \textbf{80} & \textbf{75} | \textbf{77.50} & \textbf{100} | \textbf{100} & \textbf{100} | \textbf{100} & \textbf{100} | \textbf{100} & \textbf{100} | \textbf{100} \\
Gemini-3 Pro & \textbf{100} | \textbf{100} & \textbf{100} | \textbf{100} & 40 | 60 & 40 | 44 & 60 | \textbf{80} & 60 | 66.67 & 80 | 80 & 80 | 80 & \textbf{100} | \textbf{100} & \textbf{100} | \textbf{100} \\
Gemini-3 Flash & \textbf{100} | \textbf{100} & \textbf{100} | \textbf{100} & 25 | 60 & 25 | 35 & 50 | 66.67 & 50 | 66.67 & \textbf{100} | \textbf{100} & \textbf{100} | \textbf{100} & 25 | 50 & 25 | 50 \\
Composer & \textbf{100} | \textbf{100} & \textbf{100} | \textbf{100} & 50 | 60 & 50 | 57.14 & 75 | 75 & 50 | 75 & 66.67 | 75 & 50 | 75 & \textbf{100} | \textbf{100} & \textbf{100} | \textbf{100} \\
Grok & \textbf{100} | \textbf{100} & \textbf{100} | 81.82 & 33.33 | 50 & 33.33 | 50 & 75 | \textbf{80} & 62.50 | 60.61 & 75 | 60 & 75 | 57.14 & \textbf{100} | 75 & \textbf{100} | 75 \\
\midrule
\multicolumn{11}{l}{\textit{\textbf{Cross-domain evaluation}}} \\
\midrule
GPT-5.2 & \textbf{76.47} | \textbf{81.25} & \textbf{74.23} | 70.97 & 88.24 | 87.50 & 83.82 | 75 & \textbf{80.65} | \textbf{87.50} & \textbf{80.65} | 80 & \textbf{80.65} | \textbf{87.50} & \textbf{77.94} | 77.78 & \textbf{78.57} | \textbf{83.33} & \textbf{76.47} | 69.57 \\
Claude-4.5 Opus & 72.73 | \textbf{81.25} & 72.73 | \textbf{80} & \textbf{90} | \textbf{93.75} & \textbf{90} | \textbf{93.33} & 80 | \textbf{87.50} & 80 | \textbf{87.50} & 76.92 | 84.62 & 76.92 | 84 & 75 | 81.25 & 75 | \textbf{80} \\
Gemini-3 Pro & 61.54 | 75 & 61.54 | 65.52 & 80.77 | 87.50 & 80.77 | 83.33 & 75 | 81.25 & 75 | 76.47 & 69.23 | 81.25 & 69.23 | 72 & 70 | 80 & 70 | 73.33 \\
Gemini-3 Flash & 50 | 66.67 & 50 | 56.25 & 80 | 75 & 80 | 75 & 57.14 | 70 & 57.14 | 57.14 & 42.86 | 63.64 & 42.86 | 52 & 55.56 | 72.73 & 55.56 | 61.54 \\
Composer & 68.75 | 76.19 & 58.33 | 75 & 83.33 | 87.50 & 71.43 | 87.50 & 75 | 81.82 & 71.43 | 81.25 & 76.92 | 81.82 & 69.23 | 81.25 & 71.43 | 76.19 & 61.11 | 75 \\
Grok & 68.75 | 66.67 & 65 | 60.71 & 85 | 78.57 & 81 | 68.18 & 75 | 72.22 & 75 | 66.67 & 75 | 77.78 & 71.43 | 66.67 & 71.43 | 74.07 & 68.18 | 62.50 \\
\bottomrule
\end{tabular}
}
\end{table}

\subsubsection{Representative Claude-4.5 Opus FSTab}

Table~\ref{table:fstab_claude} contains the Claude-specific FSTab that was removed from the main paper.

\begin{table}[h]
\raggedright
\small
\caption{Representative model-specific FSTab for \textbf{Claude-4.5 Opus}.}
\label{table:fstab_claude}
\resizebox{\textwidth}{!}{%
\begin{tabular}{llll}
\toprule
\textbf{Feature} & \textbf{Top Recurring Vulnerabilities} & \textbf{Feature} & \textbf{Top Recurring Vulnerabilities} \\ \midrule
Access Admin Panel & \texttt{js/missing-rate-limiting} & Manage User Permissions & \texttt{js/missing-rate-limiting} \\
Apply Search Filter & \texttt{py/flask-debug}, \texttt{js/missing-rate-limiting} & Register New Account & \texttt{js/clear-text-storage...}, \texttt{js/missing-rate-limiting} \\
Browse Product Catalog & \texttt{js/missing-rate-limiting} & Reset Forgotten Password & \texttt{js/missing-rate-limiting} \\
Create Backup & \texttt{py/path-injection} & Save New Record To DB & \texttt{py/sql-injection}, \texttt{py/stack-trace-exposure} \\
Delete Record From DB & \texttt{py/stack-trace-exposure}, \texttt{js/missing-rate-limiting} & Update Record In DB & \texttt{js/missing-rate-limiting} \\
Download File & \texttt{py/path-injection}, \texttt{py/flask-debug} & Upload Document or File & \texttt{py/stack-trace-exposure}, \texttt{js/xss-through-dom} \\
Fetch Data From DB & \texttt{js/xss-through-dom}, \texttt{py/stack-trace-exposure} & User Login (Password) & \texttt{js/remote-prop-injection}, \texttt{js/insecure-random} \\
Follow User & \texttt{js/regex/missing-regexp-anchor} & User Login (Social) & \texttt{js/missing-rate-limiting} \\
Gen/Validate Auth Token & \texttt{js/missing-rate-limiting} & View Analytics Dashboard & \texttt{js/missing-rate-limiting} \\
Like Or Upvote Content & \texttt{js/missing-rate-limiting} & View Order Status & \texttt{js/missing-rate-limiting} \\
Load Next Page & \texttt{js/regex/missing-anchor}, \texttt{js/missing-rate-limiting} & View User Profile & \texttt{js/missing-rate-limiting} \\ \bottomrule
\end{tabular}%
}
\end{table}

\subsubsection{FSTab: Model-Specific Vulnerability Databases}
\label{appendix:model_tables_filtered}

This subsubsection provides the remaining model-specific FSTab lookup mappings for the evaluated production models. Together with Table~\ref{table:fstab_claude}, these tables summarize the feature-conditioned vulnerability databases used in our attack analysis. To ensure conciseness, we display only features where at least one recurring vulnerability pattern was identified during construction.

% --- TABLE 1: GPT-5.2 ---
\begin{table}[H]
\centering
\caption{FSTab: Recurring Vulnerability Fingerprints for \textbf{GPT-5.2}.}
\label{table:fstab_gpt}
\begin{tabular}{ll}
\toprule
\textbf{feature} & \textbf{Top Recurring Vulnerabilities (Rule IDs)} \\ \midrule
Access Admin Panel & \texttt{js/missing-rate-limiting} \\
Add Item To Shopping Cart & \texttt{js/remote-property-injection}, \texttt{py/url-redirection} \\
Apply Search Filter & \texttt{js/missing-origin-check}, \texttt{js/missing-rate-limiting} \\
Delete Record From Database & \texttt{py/url-redirection} \\
Download File & \texttt{js/remote-property-injection}, \texttt{py/path-injection} \\
Fetch Data From Database & \texttt{py/path-injection} \\
Import Data From File & \texttt{py/sql-injection} \\
Load Next Page & \texttt{js/regex/missing-regexp-anchor}, \texttt{py/sql-injection} \\
Manage User Permissions & \texttt{py/url-redirection} \\
Publish New Post & \texttt{js/missing-rate-limiting} \\
Register New Account & \texttt{js/prototype-pollution-utility}, \texttt{js/missing-rate-limiting} \\
Submit Payment & \texttt{py/url-redirection} \\
Update Record In Database & \texttt{py/url-redirection}, \texttt{js/missing-rate-limiting} \\
Upload Document Or File & \texttt{py/path-injection} \\
User Login With Password & \texttt{js/client-side-unvalidated-url-redirection}, \texttt{js/xss} \\
View Inbox Messages & \texttt{js/missing-rate-limiting} \\
View User Profile & \texttt{js/missing-rate-limiting} \\ \bottomrule
\end{tabular}
\end{table}

% --- TABLE 3: Gemini-3 Pro ---
\begin{table}[H]
\centering
\caption{FSTab: Recurring Vulnerability Fingerprints for \textbf{Gemini-3 Pro}.}
\label{table:fstab_gemini_pro}
\begin{tabular}{ll}
\toprule
\textbf{feature} & \textbf{Top Recurring Vulnerabilities (Rule IDs)} \\ \midrule
Add Item To Shopping Cart & \texttt{py/flask-debug} \\
Apply Search Filter & \texttt{py/flask-debug} \\
Block User & \texttt{js/xss-through-dom} \\
Create Backup & \texttt{py/path-injection} \\
Manage User Permissions & \texttt{py/flask-debug} \\
Register New Account & \texttt{py/flask-debug} \\
Save New Record To Database & \texttt{py/url-redirection} \\
User Login With Password & \texttt{js/xss}, \texttt{js/client-side-unvalidated-url-redirection} \\
User Logout & \texttt{py/flask-debug} \\ \bottomrule
\end{tabular}
\end{table}

% --- TABLE 4: Gemini-3 Flash ---
\begin{table}[H]
\centering
\caption{FSTab: Recurring Vulnerability Fingerprints for \textbf{Gemini-3 Flash}.}
\label{table:fstab_gemini_flash}
\begin{tabular}{ll}
\toprule
\textbf{feature} & \textbf{Top Recurring Vulnerabilities (Rule IDs)} \\ \midrule
Access Admin Panel & \texttt{js/missing-rate-limiting} \\
Apply Search Filter & \texttt{py/flask-debug} \\
Generate Or Validate Auth Token & \texttt{js/missing-rate-limiting} \\
Register New Account & \texttt{js/missing-rate-limiting} \\
Upload Document Or File & \texttt{py/stack-trace-exposure}, \texttt{py/url-redirection} \\
User Login With Password & \texttt{py/flask-debug}, \texttt{js/missing-rate-limiting} \\
View User Profile & \texttt{js/missing-rate-limiting} \\ \bottomrule
\end{tabular}
\end{table}

% --- TABLE 5: Composer ---
\begin{table}[H]
\centering
\caption{FSTab: Recurring Vulnerability Fingerprints for \textbf{Composer}.}
\label{table:fstab_composer}
\begin{tabular}{ll}
\toprule
\textbf{feature} & \textbf{Top Recurring Vulnerabilities (Rule IDs)} \\ \midrule
Access Admin Panel & \texttt{js/missing-rate-limiting} \\
Apply Search Filter & \texttt{py/flask-debug}, \texttt{py/sql-injection} \\
Block User & \texttt{js/xss-through-dom} \\
Create Backup & \texttt{py/path-injection}, \texttt{py/stack-trace-exposure} \\
Delete Record From Database & \texttt{py/path-injection}, \texttt{py/stack-trace-exposure} \\
Download File & \texttt{py/path-injection}, \texttt{py/stack-trace-exposure} \\
Fetch Data From Database & \texttt{js/request-forgery}, \texttt{py/stack-trace-exposure} \\
Generate Or Validate Auth Token & \texttt{js/missing-rate-limiting} \\
Like Or Upvote Content & \texttt{js/missing-rate-limiting} \\
Load Next Page & \texttt{js/missing-rate-limiting} \\
Manage User Permissions & \texttt{js/missing-rate-limiting} \\
Post Comment & \texttt{js/missing-rate-limiting} \\
Register New Account & \texttt{js/sql-injection}, \texttt{js/missing-rate-limiting} \\
Save New Record To Database & \texttt{py/stack-trace-exposure}, \texttt{js/missing-rate-limiting} \\
Search Content & \texttt{js/missing-rate-limiting} \\
Send Invitation & \texttt{js/missing-rate-limiting} \\
Share Content & \texttt{js/missing-rate-limiting} \\
Sort Results & \texttt{js/sql-injection}, \texttt{py/flask-debug} \\
Submit Payment & \texttt{js/missing-rate-limiting} \\
Update Record In Database & \texttt{js/missing-rate-limiting} \\
Upload Document Or File & \texttt{py/stack-trace-exposure}, \texttt{js/missing-rate-limiting} \\
User Login With Password & \texttt{js/prototype-polluting-assignment}, \texttt{js/remote-property-injection} \\
User Login With Social Account & \texttt{js/missing-rate-limiting} \\
View Analytics Dashboard & \texttt{js/missing-rate-limiting} \\
View Inbox Messages & \texttt{js/missing-rate-limiting} \\
View Order Status & \texttt{js/sql-injection}, \texttt{js/missing-rate-limiting} \\
View Shopping Cart & \texttt{js/missing-rate-limiting} \\
View User Profile & \texttt{js/missing-rate-limiting} \\ \bottomrule
\end{tabular}
\end{table}

% --- TABLE 6: Grok ---
\begin{table}[H]
\centering
\caption{FSTab: Recurring Vulnerability Fingerprints for \textbf{Grok}.}
\label{table:fstab_grok}
\begin{tabular}{ll}
\toprule
\textbf{feature} & \textbf{Top Recurring Vulnerabilities (Rule IDs)} \\ \midrule
Access Admin Panel & \texttt{py/stack-trace-exposure} \\
Apply Search Filter & \texttt{py/flask-debug}, \texttt{py/stack-trace-exposure} \\
Block User & \texttt{js/functionality-from-untrusted-source} \\
Create Backup & \texttt{py/log-injection}, \texttt{py/stack-trace-exposure} \\
Delete Record From Database & \texttt{py/path-injection}, \texttt{py/stack-trace-exposure} \\
Download File & \texttt{py/path-injection} \\
Fetch Data From Database & \texttt{py/sql-injection}, \texttt{js/xss-through-dom} \\
Generate Or Validate Auth Token & \texttt{js/sql-injection}, \texttt{js/missing-rate-limiting} \\
Like Or Upvote Content & \texttt{js/missing-rate-limiting} \\
Load Next Page & \texttt{js/remote-property-injection}, \texttt{js/sql-injection} \\
Proceed To Checkout & \texttt{py/flask-debug} \\
Register New Account & \texttt{py/flask-debug}, \texttt{js/missing-rate-limiting} \\
Save New Record To Database & \texttt{js/sql-injection}, \texttt{py/stack-trace-exposure} \\
Search Content & \texttt{js/sql-injection} \\
Sort Results & \texttt{js/missing-rate-limiting} \\
Submit Payment & \texttt{js/missing-rate-limiting} \\
Update Record In Database & \texttt{js/sql-injection}, \texttt{py/stack-trace-exposure} \\
Upload Document Or File & \texttt{py/url-redirection}, \texttt{py/path-injection} \\
User Login With Password & \texttt{js/redos}, \texttt{js/clear-text-storage-of-sensitive-data} \\
View Analytics Dashboard & \texttt{js/missing-rate-limiting}, \texttt{py/stack-trace-exposure} \\
View Inbox Messages & \texttt{js/missing-rate-limiting} \\
View Order Status & \texttt{js/missing-rate-limiting} \\
View Shopping Cart & \texttt{js/missing-rate-limiting} \\
View User Profile & \texttt{js/missing-rate-limiting}, \texttt{js/sql-injection} \\ \bottomrule
\end{tabular}
\end{table}

\clearpage

\subsection{Tables}
\label{appendix:framework_tables}

\subsubsection{Construction-Budget Ablation}
\label{app:budget_ablation}

\begin{figure}[H]
    \centering
    \includegraphics[width=0.8\linewidth]{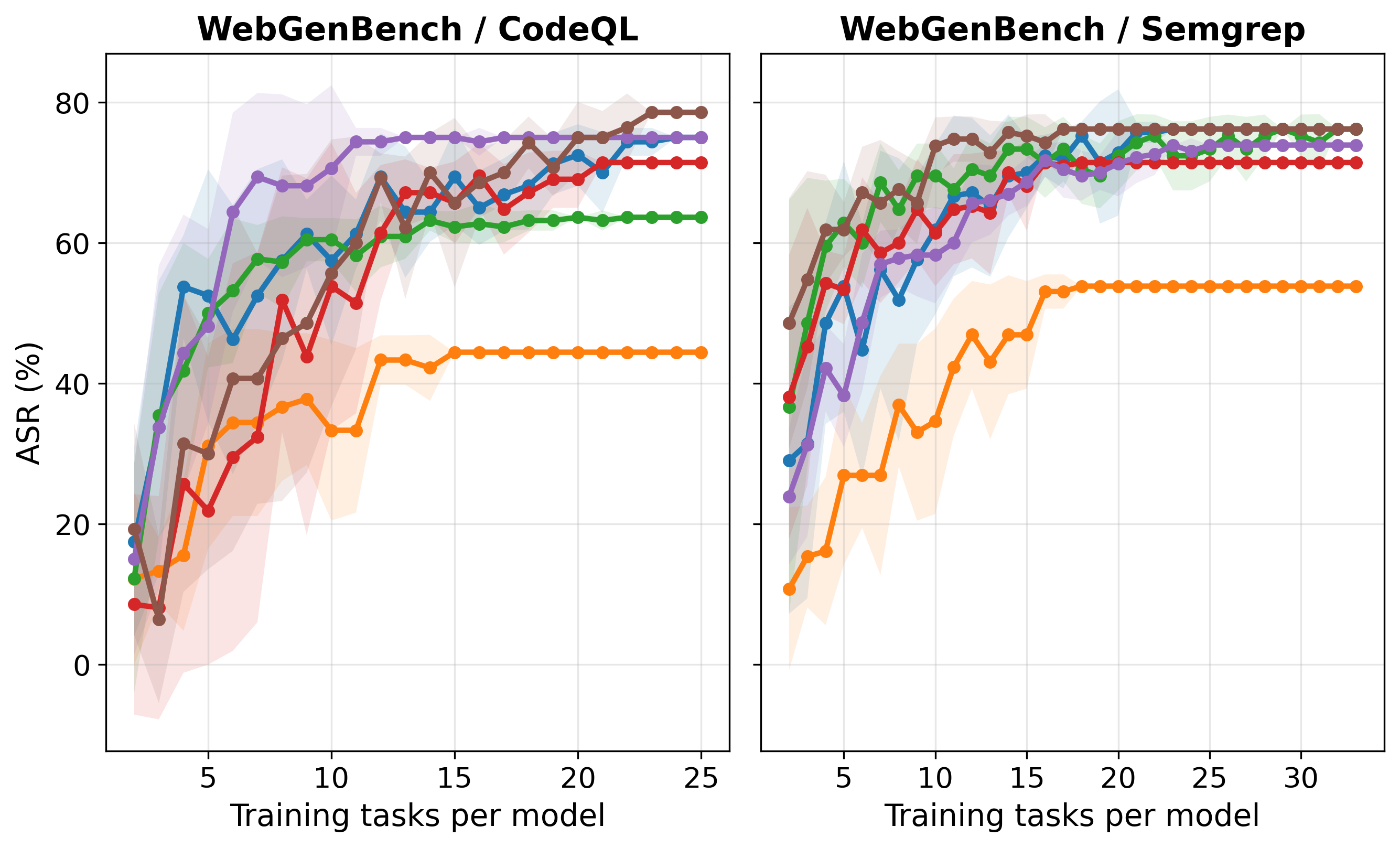}
    \caption{ASR vs.\ construction budget (WebGenBench, FSTab with $N{=}1$). Both benchmarks plateau around 20 projects.}
    \label{fig:ablation_asr}
\end{figure}

\begin{figure}[H]
    \centering
    \includegraphics[width=0.8\linewidth]{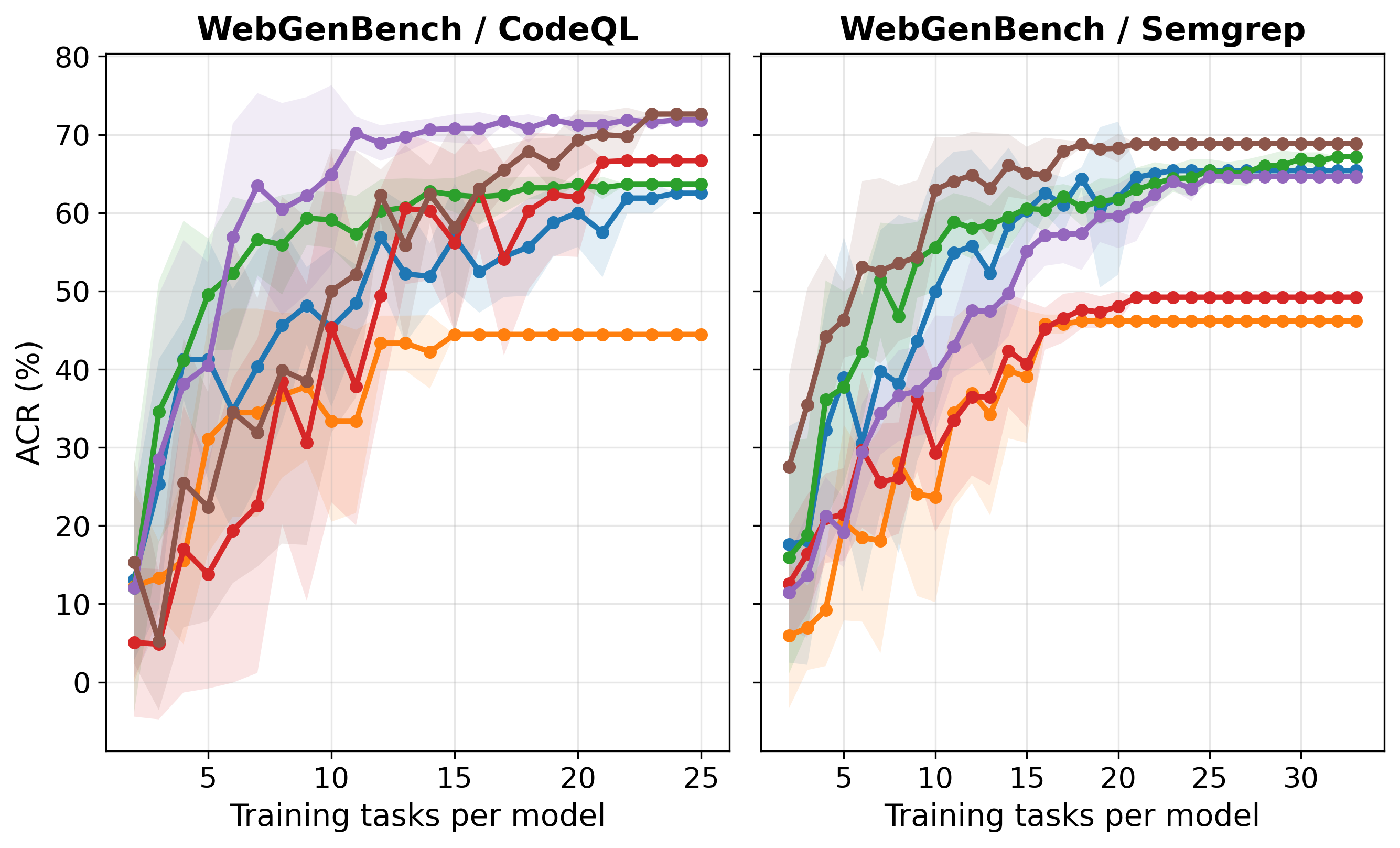}
    \includegraphics[width=0.8\linewidth]{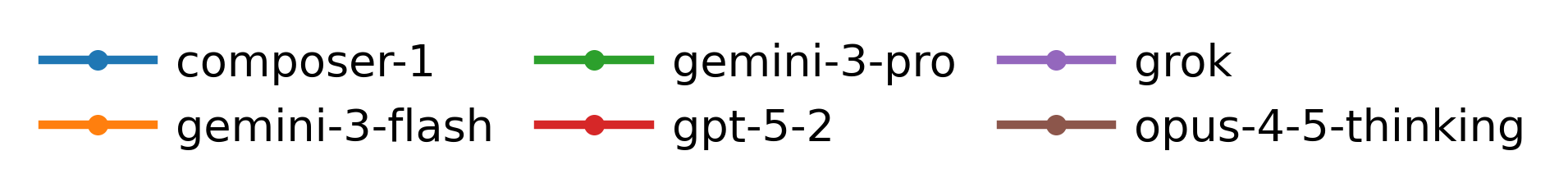}
    \caption{ACR vs.\ construction budget under the same sweep. The plateau near 20 projects motivates our construction budget choice.}
    \label{fig:ablation_acr}
\end{figure}

The construction-budget ablation was removed from the main paper for space. Figure~\ref{fig:ablation_asr} and Figure~\ref{fig:ablation_acr} show that both ASR and ACR plateau around 20 construction projects, which motivates our default budget choice.

\subsubsection{Attack Rule Budget and CWE Coverage}
\label{app:rule_budget_cwe}

Because FSTab relies on a minimum-support threshold to filter noise, it learns only a tightly bounded space of the most recurring vulnerabilities. In our construction pipeline, we retain only feature--rule pairs observed at least three times ($C(f,r)\geq 3$). This yields 28 learnable CodeQL rules and 39 learnable Semgrep rules, even though the underlying scanners expose 300+ and 5,000+ rules, respectively. After deduplicating repeated rule IDs across the feature-conditioned query results for a target program, the average prediction set remains small: 4.86 CodeQL rules per project and 8.03 Semgrep rules per project. The resulting coverage still spans critical vulnerabilities families such as CWE-89 (SQL Injection), CWE-79 (Cross-Site Scripting), CWE-22 (Path Traversal), and CWE-20 (Improper Input Validation).

\begin{table}[H]
\centering
\caption{R6: CWE coverage summary for the bounded FSTab rule budget.}
\label{tab:r6_cwe_coverage}
\begin{tabular}{lcc}
\toprule
\textbf{Statistic} & \textbf{CodeQL} & \textbf{Semgrep} \\
\midrule
Learnable rules & 28 & 39 \\
Distinct CWE IDs & 40 & 22 \\
Total scanner rules & 300+ & 5,000+ \\
Avg. rules/project & 4.86 & 8.03 \\
\bottomrule
\end{tabular}
\end{table}

This bounded budget matters operationally. An external auditor can realistically inspect 5-8 high-priority rules per project; triaging thousands of generic scanner alerts is not feasible. The results therefore show that recurrence is not only statistically measurable, but practically usable.

Representative CodeQL coverage includes CWE-89 (SQL Injection), CWE-79 (Cross-Site Scripting), CWE-770 (Missing Rate Limiting), CWE-601 (Open Redirect), CWE-22/CWE-23/CWE-73 (Path Traversal variants), CWE-338 (Weak Randomness), CWE-730/CWE-1333 (ReDoS), CWE-312/CWE-315 (Cleartext Storage), CWE-209 (Stack Trace Exposure), CWE-918 (SSRF), CWE-807 (Authentication Bypass), CWE-94 (Code Injection), CWE-400 (Resource Exhaustion), CWE-489 (Debug Mode), CWE-434 (Unrestricted Upload), and CWE-307 (Brute Force), among others.

Representative Semgrep coverage includes CWE-352 (CSRF), CWE-522/CWE-798 (Hardcoded Credentials), CWE-79 (XSS), CWE-89 (SQL Injection), CWE-489 (Debug Mode), CWE-521 (Weak Password), CWE-601 (Open Redirect), CWE-942 (CORS Misconfiguration), CWE-502 (Deserialization), CWE-319 (Cleartext Transmission), CWE-1333 (ReDoS), CWE-22 (Path Traversal), and CWE-250 (Unnecessary Privileges), among others.

\subsubsection{Exhaustive Feature Level Vulnerability Recurrence (FVR)}
\label{appendix:FVR_tables}

This subsubsection provides the exhaustive feature-level Vulnerability Recurrence (FVR) scores for each of the six evaluated production models. The FVR score represents the stability of vulnerability patterns for a given feature. Values closer to 100 indicate high persistence in the model's security failures for that specific feature.

% --- TABLE 1: USER LIFECYCLE & AUTHENTICATION ---
\begin{table}[H]
\centering
\caption{FVR Scores: Active User Lifecycle and Security Triggers.}
\begin{tabular}{l cccc}
\toprule
\textbf{Model} & \textbf{\shortstack{Register\\Account}} $\uparrow$ & \textbf{\shortstack{User Login\\(Password)}} $\uparrow$ & \textbf{\shortstack{User\\Logout}} $\uparrow$ & \textbf{\shortstack{Generate\\Auth Token}} $\uparrow$ \\ \midrule
GPT-5.2         & \textbf{100.00} & 40.91          & 0.00            & 0.00            \\
Claude-4.5 Opus & 66.67           & 52.94          & 0.00            & 50.00           \\
Gemini-3 Pro    & \textbf{100.00} & \textbf{71.43} & \textbf{100.00} & 0.00            \\
Gemini-3 Flash  & \textbf{100.00} & 60.00          & 0.00            & \textbf{100.00} \\
Composer        & 40.00           & 60.87          & 0.00            & 50.00           \\
Grok            & \textbf{100.00} & 68.42          & 0.00            & \textbf{100.00} \\ \bottomrule
\end{tabular}
\end{table}

% --- TABLE 2: DATABASE & DATA OPERATIONS ---
\begin{table}[H]
\centering
\caption{FVR Scores: Active Database and Persistence Triggers.}
\begin{tabular}{l ccccc}
\toprule
\textbf{Model} & \textbf{\shortstack{Delete\\Record}} $\uparrow$ & \textbf{\shortstack{Save New\\Record}} $\uparrow$ & \textbf{\shortstack{Update\\Record}} $\uparrow$ & \textbf{\shortstack{Fetch\\Data}} $\uparrow$ & \textbf{\shortstack{Create\\Backup}} $\uparrow$ \\ \midrule
GPT-5.2         & \textbf{100.00} & 25.00          & 40.00          & 20.00          & 0.00 \\
Claude-4.5 Opus & 40.00           & 57.14          & 25.00          & 33.33          & 0.00 \\
Gemini-3 Pro    & 0.00            & \textbf{66.67} & 0.00           & 0.00           & 0.00 \\
Gemini-3 Flash  & 0.00            & 0.00           & 0.00           & 0.00           & 0.00 \\
Composer        & \textbf{100.00} & 40.00          & 50.00          & \textbf{50.00} & \textbf{33.33} \\
Grok            & 25.00           & 28.57          & \textbf{66.67} & 33.33          & 20.00 \\ \bottomrule
\end{tabular}
\end{table}

% --- TABLE 3: UI & NAVIGATION ---
\begin{table}[H]
\centering
\caption{FVR Scores: Active UI and Search Interaction Triggers.}
\begin{tabular}{l ccccc}
\toprule
\textbf{Model} & \textbf{\shortstack{Handle\\UI Event}} $\uparrow$ & \textbf{\shortstack{Toggle\\UI Elem.}} $\uparrow$ & \textbf{\shortstack{Load\\Next Pg}} $\uparrow$ & \textbf{\shortstack{Apply\\Filter}} $\uparrow$ & \textbf{\shortstack{Sort\\Results}} $\uparrow$ \\ \midrule
GPT-5.2         & \textbf{100.00} & 0.00           & \textbf{75.00} & 40.00          & \textbf{100.00} \\
Claude-4.5 Opus & 0.00            & 0.00           & 50.00          & 60.00          & 33.33           \\
Gemini-3 Pro    & 0.00            & 0.00           & 50.00          & \textbf{100.00} & \textbf{100.00} \\
Gemini-3 Flash  & 0.00            & 0.00           & 0.00           & 20.00          & 0.00            \\
Composer        & 66.67           & \textbf{66.67} & 40.00          & \textbf{75.00} & \textbf{100.00} \\
Grok            & 0.00            & 0.00           & 50.00          & \textbf{80.00} & \textbf{100.00} \\ \bottomrule
\end{tabular}
\end{table}

% --- TABLE 4: INTEGRATION & LIFECYCLE ---
\begin{table}[H]
\centering
\caption{FVR Scores: Active System Integration and Middleware Triggers.}
\begin{tabular}{l ccccc}
\toprule
\textbf{Model} & \textbf{\shortstack{API\\Request}} $\uparrow$ & \textbf{\shortstack{Websocket\\Conn.}} $\uparrow$ & \textbf{\shortstack{Process\\Mid.ware}} $\uparrow$ & \textbf{\shortstack{Init.\\App}} $\uparrow$ & \textbf{\shortstack{Run\\Test}} $\uparrow$ \\ \midrule
GPT-5.2         & 66.67           & 0.00            & 0.00            & 33.33          & 0.00 \\
Claude-4.5 Opus & \textbf{75.00}  & 33.33           & 0.00            & 37.50          & 50.00 \\
Gemini-3 Pro    & \textbf{100.00} & 50.00           & 0.00            & 50.00          & 50.00 \\
Gemini-3 Flash  & 50.00           & 0.00            & 0.00            & \textbf{66.67} & \textbf{100.00} \\
Composer        & \textbf{80.00}  & \textbf{100.00} & \textbf{100.00} & 40.00          & 0.00 \\
Grok            & 50.00           & 60.00           & \textbf{100.00} & 50.00          & 0.00 \\ \bottomrule
\end{tabular}
\end{table}

% --- TABLE 5: SOCIAL & PUBLISHING ---
\begin{table}[H]
\centering
\caption{FVR Scores: Active Social Engagement and Publishing Triggers.}
\begin{tabular}{l ccccc}
\toprule
\textbf{Model} & \textbf{\shortstack{Like/Upvote\\Content}} $\uparrow$ & \textbf{\shortstack{Follow\\User}} $\uparrow$ & \textbf{\shortstack{Block\\User}} $\uparrow$ & \textbf{\shortstack{Publish\\New Post}} $\uparrow$ & \textbf{\shortstack{Edit\\Exist. Post}} $\uparrow$ \\ \midrule
GPT-5.2         & 0.00            & 0.00 & 0.00           & \textbf{100.00} & 0.00 \\
Claude-4.5 Opus & \textbf{100.00} & 50.00 & 0.00           & 0.00            & 0.00 \\
Gemini-3 Pro    & \textbf{100.00} & 0.00 & \textbf{100.00} & \textbf{100.00} & \textbf{100.00} \\
Gemini-3 Flash  & \textbf{100.00} & 0.00 & 0.00           & \textbf{100.00} & 0.00 \\
Composer        & 33.33           & 0.00 & 50.00          & 0.00            & 0.00 \\
Grok            & \textbf{100.00} & 0.00 & 33.33          & 0.00            & 0.00 \\ \bottomrule
\end{tabular}
\end{table}

% --- TABLE 6: FILE HANDLING & TRANSFERS ---
\begin{table}[H]
\centering
\caption{FVR Scores: Active File Handling and Download Triggers.}
\begin{tabular}{l ccc}
\toprule
\textbf{Model} & \textbf{\shortstack{Download\\File}} $\uparrow$ & \textbf{\shortstack{Upload\\Doc/File}} $\uparrow$ & \textbf{\shortstack{Share\\Content}} $\uparrow$ \\ \midrule
GPT-5.2         & 60.00           & 50.00 & 0.00 \\
Claude-4.5 Opus & \textbf{100.00} & 50.00 & \textbf{50.00} \\
Gemini-3 Pro    & 0.00            & 50.00 & 0.00 \\
Gemini-3 Flash  & 0.00            & 33.33 & 0.00 \\
Composer        & \textbf{80.00}  & \textbf{66.67} & \textbf{50.00} \\
Grok            & 50.00           & 37.50 & 0.00 \\ \bottomrule
\end{tabular}
\end{table}

% --- TABLE 7: PLATFORM ADMINISTRATION ---
\begin{table}[H]
\centering
\caption{FVR Scores: Active Administrative and Platform Control Triggers.}
\begin{tabular}{l cccc}
\toprule
\textbf{Model} & \textbf{\shortstack{Admin\\Panel}} $\uparrow$ & \textbf{\shortstack{User\\Perms}} $\uparrow$ & \textbf{\shortstack{Manage\\Users}} $\uparrow$ & \textbf{\shortstack{Analytics\\Dash}} $\uparrow$ \\ \midrule
GPT-5.2         & 50.00           & 0.00            & 0.00            & 0.00 \\
Claude-4.5 Opus & \textbf{100.00} & \textbf{100.00} & 0.00            & 0.00 \\
Gemini-3 Pro    & 0.00            & \textbf{100.00} & 0.00            & 0.00 \\
Gemini-3 Flash  & \textbf{100.00} & 0.00            & 0.00            & 0.00 \\
Composer        & 50.00           & \textbf{100.00} & 0.00            & 0.00 \\
Grok            & 0.00            & 0.00            & \textbf{100.00} & 50.00 \\ \bottomrule
\end{tabular}
\end{table}

% --- TABLE 8: VIEWS & COMMERCE FUNNEL ---
\begin{table}[H]
\centering
\caption{FVR Scores: Active Content Views and Transactional Triggers.}
\begin{tabular}{l ccccc}
\toprule
\textbf{Model} & \textbf{\shortstack{View Inbox\\Messages}} $\uparrow$ & \textbf{\shortstack{View Order\\Status}} $\uparrow$ & \textbf{\shortstack{View Shopping\\Cart}} $\uparrow$ & \textbf{\shortstack{View\\Calendar}} $\uparrow$ & \textbf{\shortstack{Add To\\Cart}} $\uparrow$ \\ \midrule
GPT-5.2         & \textbf{100.00} & 0.00 & 0.00 & 0.00 & 25.00 \\
Claude-4.5 Opus & 0.00            & \textbf{100.00} & 0.00 & \textbf{100.00} & 0.00 \\
Gemini-3 Pro    & 0.00            & 0.00 & 0.00 & 0.00 & \textbf{100.00} \\
Gemini-3 Flash  & 0.00            & 0.00 & 0.00 & 0.00 & 0.00 \\
Composer        & \textbf{100.00} & \textbf{100.00} & \textbf{100.00} & \textbf{100.00} & 0.00 \\
Grok            & 50.00           & 0.00 & 0.00 & 50.00 & 0.00 \\ \bottomrule
\end{tabular}
\end{table}

% --- TABLE 9: COMMERCE WORKFLOW & GENERIC UTILITIES ---
\begin{table}[H]
\centering
\caption{FVR Scores: Active Commerce Workflows and Multi-Agent Utilities.}
\begin{tabular}{l ccccc}
\toprule
\textbf{Model} & \textbf{\shortstack{Proceed To\\Checkout}} $\uparrow$ & \textbf{\shortstack{Submit\\Payment}} $\uparrow$ & \textbf{\shortstack{Login\\Generic}} $\uparrow$ & \textbf{\shortstack{Post\\Comment}} $\uparrow$ & \textbf{\shortstack{Utility\\Helper}} $\uparrow$ \\ \midrule
GPT-5.2         & 0.00 & 0.00 & \textbf{100.00} & 0.00 & 0.00 \\
Claude-4.5 Opus & 0.00 & 0.00 & 50.00 & 0.00 & 0.00 \\
Gemini-3 Pro    & 0.00 & 0.00 & \textbf{100.00} & 0.00 & 0.00 \\
Gemini-3 Flash  & 0.00 & 0.00 & 0.00 & 0.00 & 0.00 \\
Composer        & 0.00 & 25.00 & 33.33 & \textbf{50.00} & \textbf{50.00} \\
Grok            & \textbf{50.00} & 33.33 & 66.67 & 0.00 & 0.00 \\ \bottomrule
\end{tabular}
\end{table}

\subsubsection{Exhaustive Rephrasing Vulnerability Persistence (RVP)}
\label{appendix:RVP_tables}

This subsection provides a prompt-slot diagnostic view of the rephrasing experiment. The formal aggregate metric $RVP_{\text{model}}$ is defined in Section~\ref{sec:rvp} and averages recurrence jointly across all five realizations of each task. Table~\ref{tab:rvp_scores}, by contrast, breaks the same experiment out by prompt slot: P0 denotes the original prompt, and P1--P4 denote the four semantic-preserving rewrites. The goal is to show whether persistence is roughly uniform across prompt realizations or disproportionately driven by particular prompt slots, rather than to redefine RVP itself.

\begin{table}[H]
\centering
\caption{RVP Scores across Prompt Variations (CodeQL Only, \%)}
\label{tab:rvp_scores}
\begin{tabular}{l ccccc}
\toprule
\textbf{Model} & \textbf{P0} $\uparrow$ & \textbf{P1} $\uparrow$ & \textbf{P2} $\uparrow$ & \textbf{P3} $\uparrow$ & \textbf{P4} $\uparrow$ \\
\midrule
GPT-5.2 & 49.70 & 45.00 & 23.94 & 39.50 & 42.86 \\
Claude-4.5 Opus & 32.95 & \textbf{66.96} & 54.28 & 32.98 & 37.69 \\
Gemini-3 Pro & 31.16 & 44.93 & 53.79 & 40.91 & 53.57 \\
Gemini-3 Flash & 31.67 & 23.70 & \textbf{83.33} & 47.50 & 55.00 \\
Composer & \textbf{49.98} & 53.29 & 54.92 & \textbf{52.73} & \textbf{56.43} \\
Grok & 23.68 & 30.37 & 24.91 & 17.48 & 36.82 \\
\bottomrule
\end{tabular}
\end{table}

\subsubsection{Exhaustive Domain Recurrence and Transfer (DVR \& CDT)}
\label{appendix:DVR_CDT_tables}

The following tables present the Within-Domain Recurrence (DVR) and Cross-Domain Transfer (CDT) scores. High CDT scores indicate that vulnerability patterns generalize across disparate application domains.

% --- TABLE: DVR ---
\begin{table}[H]
\centering
\caption{Domain Vulnerability Recurrence (DVR) Scores (CodeQL Only, \%)}
\label{tab:dvr_scores}
\begin{tabular}{l ccccc}
\toprule
\textbf{Model} & \textbf{\shortstack{E-Commerce}} $\uparrow$ & \textbf{\shortstack{Publishing}} $\uparrow$ & \textbf{\shortstack{Social}} $\uparrow$ & \textbf{\shortstack{Analytics}} $\uparrow$ & \textbf{\shortstack{Internal}} $\uparrow$ \\
\midrule
GPT-5.2 & 46.67 & 21.21 & 34.62 & 28.00 & 39.13 \\
Claude-4.5 Opus & 48.78 & 37.50 & 13.89 & 30.00 & 28.57 \\
Gemini-3 Pro & 33.33 & 47.06 & 44.44 & 40.00 & 42.11 \\
Gemini-3 Flash & 37.50 & 26.32 & \textbf{50.00} & 25.00 & 41.67 \\
Composer & \textbf{50.94} & \textbf{50.00} & 44.23 & \textbf{43.59} & \textbf{43.40} \\
Grok & 37.04 & 25.00 & 29.55 & 25.45 & 22.22 \\
\bottomrule
\end{tabular}
\end{table}

% --- TABLE: CDT ---
\begin{table}[H]
\centering
\caption{Cross-Domain Vulnerability Transfer (CDT) Scores (CodeQL Only, \%)}
\label{tab:cdt_scores}
\begin{tabular}{l ccccc}
\toprule
\textbf{Model} & \textbf{\shortstack{E-Commerce}} $\uparrow$ & \textbf{\shortstack{Publishing}} $\uparrow$ & \textbf{\shortstack{Social}} $\uparrow$ & \textbf{\shortstack{Analytics}} $\uparrow$ & \textbf{\shortstack{Internal}} $\uparrow$ \\
\midrule
GPT-5.2 & 73.33 & 39.39 & 26.92 & 24.00 & 47.83 \\
Claude-4.5 Opus & 53.66 & 55.00 & 41.67 & \textbf{64.52} & 53.06 \\
Gemini-3 Pro & 60.00 & \textbf{82.35} & 55.56 & 53.33 & 42.11 \\
Gemini-3 Flash & \textbf{87.50} & 47.37 & 37.50 & 33.33 & \textbf{58.33} \\
Composer & 56.60 & 70.59 & 50.00 & 64.10 & 45.28 \\
Grok & 55.56 & 62.50 & \textbf{59.09} & 57.14 & 52.17 \\
\bottomrule
\end{tabular}
\end{table}

\clearpage

\section{E2EDev Attack Performance}
\label{app:e2edev_appendix}
Table~\ref{tab:attack_e2edev_appendix} presents the detailed ASR and ACR results for the E2EDev dataset across all models and domains, where ASR denotes at-least-one overlap with the scanner findings. Values are presented as CodeQL / Semgrep.
Relative to WebGenBench, the performance drop on E2EDev is driven primarily by feature sparsity rather than by a contradiction of the recurrence hypothesis. E2EDev prompts intentionally describe smaller applications, and the resulting projects expose only 2.5 extracted UI features per project on average, compared with 8.6 on WebGenBench. Because FSTab is trained on WebGenBench-style programs, E2EDev constitutes a genuine out-of-distribution transfer setting.

\begin{table}[!htbp]
\centering
\caption{Overlap-based prediction performance on the E2EDev dataset. We report CodeQL and Semgrep results for ASR and ACR (in \%), where ASR denotes at-least-one overlap with the scanner findings. Values are presented as CodeQL / Semgrep.}
\label{tab:attack_e2edev_appendix}
\resizebox{\textwidth}{!}{
\begin{tabular}{lcccccccccc}
\toprule
\textbf{Model} &
\multicolumn{2}{c}{\textbf{E-commerce}} &
\multicolumn{2}{c}{\textbf{Internal Tools}} &
\multicolumn{2}{c}{\textbf{Social Media}} &
\multicolumn{2}{c}{\textbf{Blogging}} &
\multicolumn{2}{c}{\textbf{Dashboards}} \\
& ASR $\uparrow$ & ACR $\uparrow$ & ASR $\uparrow$ & ACR $\uparrow$ & ASR $\uparrow$ & ACR $\uparrow$ & ASR $\uparrow$ & ACR $\uparrow$ & ASR $\uparrow$ & ACR $\uparrow$ \\
\midrule
\multicolumn{11}{l}{\textit{\textbf{Held-out (target-domain) evaluation}}} \\
\midrule
GPT-5.2 & -- / 0 & -- / 0 & \textbf{100} / 33.33 & \textbf{100} / 33.33 & 0 / \textbf{100} & 0 / \textbf{100} & \textbf{100} / 75 & \textbf{100} / 75 & \textbf{100} / 50 & \textbf{100} / 50 \\
Claude-4.5 Opus & \textbf{100} / 0 & \textbf{100} / 0 & 50 / \textbf{100} & 50 / \textbf{100} & 0 / \textbf{100} & 0 / \textbf{100} & \textbf{100} / \textbf{100} & \textbf{100} / \textbf{100} & 0 / 50 & 0 / 50 \\
Gemini-3 Pro & \textbf{100} / \textbf{100} & \textbf{100} / \textbf{100} & 33.33 / 33.33 & 33.33 / 33.33 & \textbf{100} / \textbf{100} & \textbf{100} / \textbf{100} & \textbf{100} / 75 & \textbf{100} / 75 & \textbf{100} / \textbf{100} & \textbf{100} / \textbf{100} \\
Gemini-3 Flash & \textbf{100} / \textbf{100} & \textbf{100} / \textbf{100} & \textbf{100} / 50 & \textbf{100} / 50 & \textbf{100} / \textbf{100} & \textbf{100} / \textbf{100} & \textbf{100} / 75 & \textbf{100} / 75 & \textbf{100} / 66.67 & \textbf{100} / 66.67 \\
Composer & -- / 0 & -- / 0 & 50 / 50 & 50 / 50 & 0 / \textbf{100} & 0 / \textbf{100} & \textbf{100} / 50 & \textbf{100} / 50 & 0 / 0 & 0 / 0 \\
Grok & \textbf{100} / \textbf{100} & \textbf{100} / \textbf{100} & -- / 50 & -- / 50 & 0 / \textbf{100} & 0 / \textbf{100} & \textbf{100} / 50 & \textbf{100} / 50 & \textbf{100} / \textbf{100} & \textbf{100} / \textbf{100} \\
\midrule
\multicolumn{11}{l}{\textit{\textbf{Cross-domain evaluation}}} \\
\midrule
GPT-5.2 & 75 / 63.64 & 75 / 63.64 & 66.67 / 66.67 & 66.67 / 66.67 & \textbf{100} / 50 & \textbf{100} / 50 & 66.67 / 50 & 66.67 / 50 & 66.67 / 60 & 66.67 / 60 \\
Claude-4.5 Opus & 40 / \textbf{90.91} & 40 / \textbf{90.91} & 50 / 77.78 & 50 / 77.78 & 60 / \textbf{77.78} & 60 / \textbf{77.78} & 40 / \textbf{77.78} & 40 / \textbf{77.78} & 60 / \textbf{90} & 60 / \textbf{90} \\
Gemini-3 Pro & 71.43 / 72.73 & 71.43 / 72.73 & \textbf{100} / \textbf{88.89} & \textbf{100} / \textbf{88.89} & 75 / 70 & 75 / 70 & 71.43 / 75 & 71.43 / 75 & 75 / 70 & 75 / 70 \\
Gemini-3 Flash & \textbf{83.33} / 66.67 & \textbf{83.33} / 66.67 & \textbf{100} / 71.43 & \textbf{100} / 71.43 & 75 / 58.33 & 75 / 58.33 & 66.67 / 63.64 & 66.67 / 63.64 & 75 / 57.14 & 75 / 57.14 \\
Composer & 50 / 54.55 & 50 / 54.55 & 66.67 / 66.67 & 66.67 / 66.67 & 60 / 70 & 60 / 70 & 50 / 57.14 & 50 / 57.14 & 60 / 50 & 60 / 50 \\
Grok & 75 / 72.73 & 75 / 72.73 & 60 / 70 & 60 / 70 & 75 / 66.67 & 75 / 66.67 & 66.67 / 66.67 & 66.67 / 66.67 & 75 / 66.67 & 75 / 66.67 \\
\bottomrule
\end{tabular}
}
\end{table}

\section{Statistical Significance Summary}
\label{app:stats_summary}

For the held-out WebGenBench attack comparison, we report 95\% Wilson confidence intervals (closed-form binomial, with no Normality assumption) on aggregate ASR over project--model evaluation units. We do not report corresponding Wilson intervals for E2EDev. For stochastic experiments: Random-Budget sampling, wrong-fingerprint pairing, and label-noise perturbations. We report the 1$\sigma$ sample standard deviation across 10 independent random seeds.

\begin{table}[H]
\centering
\small
\caption{Aggregate held-out WebGenBench ASR with 95\% Wilson confidence intervals over project--model evaluation units. The non-overlapping intervals show a clean separation between FSTab and the matched-budget Random-Budget baseline.}
\label{tab:attack_aggregate_ci}
\begin{tabular}{lcc}
\toprule
\textbf{Eval} & \textbf{FSTab ASR} & \textbf{Random-Budget ASR} \\
\midrule
WebGen CodeQL & 69.4\% [59.7--77.6] & 9.4\% [4.4--17.4] \\
WebGen Semgrep & 72.5\% [63.9--79.7] & 20.8\% [14.5--28.6] \\
\bottomrule
\end{tabular}
\end{table}

\begin{table}[H]
\centering
\small
\caption{Seed variation for stochastic robustness experiments. Values report 1$\sigma$ sample standard deviation over 10 independent random seeds.}
\label{tab:seed_std_summary}
\begin{tabular}{lc}
\toprule
\textbf{Experiment} & \textbf{Seed std (1$\sigma$ over 10 seeds)} \\
\midrule
Random-Budget & ASR $\pm$2.3--2.6 pts \\
Wrong-fingerprint & ASR $\pm$1.6--1.9 pts \\
Label-noise (drop / add / swap) & ASR $\pm$0.6--2.3 pts \\
Rule-frequency tail-only & ASR $\pm$3.5--6.0 pts \\
\bottomrule
\end{tabular}
\end{table}

\section{Static Analysis Validation}
\label{appendix:label_validation}

To rigorously assess the fidelity of our automated ground-truth generation, we implemented a human-in-the-loop validation protocol on a randomly sampled subset of the generated projects. The objective was to quantify the rate of false positives in our static analysis pipeline and ensure that the recurrence metrics report actionable security flaws rather than tool artifacts.

\paragraph{Methodology}
We performed a manual code audit on the sampled projects, specifically verifying the exploitability and reachability of vulnerabilities flagged by our dual-engine scanner (CodeQL and Semgrep). For each flagged instance, a human expert reviewed the generated source code to determine if the detected pattern constituted a genuine violation of the associated CWE definition (True Positive) or just harmless code (False Positive).

\paragraph{Precision Analysis}
Our audit revealed precision rates of \textbf{82\%} for CodeQL and \textbf{74\%} for Semgrep. While static analysis always misses some nuance, these scores show that the majority of flagged items are actual insecure coding patterns. Furthermore, since FSTab focuses on comparing rates between models rather than counting total vulnerabilities, these validation rates suggest our metrics are conservative. The fact that both scanners found consistent patterns confirms that findings like the ``Universality Gap'' come from the models themselves, not from errors in a specific tool.

\newpage

\section{Semantic Feature Extraction Example}
\label{app:feature_extraction_example}

This section provides a concrete, end-to-end example of how our automated pipeline assigns an observable
frontend feature (a semantic UI action) to a specific code location. The example follows the procedure
in Section~\ref{semantic_action_extraction}: (i) isolate the enclosing code context around a target line and (ii) score candidate
actions using the heuristic relevance function $S(a, C_\ell)$, assigning $\hat{a}=\arg\max_a S(a, C_\ell)$.

\subsection{Example Setup: JavaScript WebSocket Chat Handler}
\label{app:feature_extraction_example:setup}

\paragraph{Input}
We consider a client-side JavaScript file in a chat plugin that supports sending messages and sharing files.
A static analyzer flags a finding at line $\ell=17$ (file-link construction in the WebSocket receive handler).
The feature extractor is given the pair (file path, line number) and returns a semantic UI action label.

\paragraph{Local code context}
The following snippet shows the relevant lines (the context window used by the extractor includes the full
\texttt{ws.onmessage} handler surrounding the target line):

\begin{verbatim}
14: } else if (data.type === 'file') {
15:     messageDiv.classList.add('received');
16:     var link = document.createElement('a');
17:     link.href = data.url;
18:     link.className = 'file-link';
19:     link.target = '_blank';
20:     link.innerHTML = 'Clip: ' + data.filename;
\end{verbatim}

\subsection{Step 1: Structural Context Extraction}
\label{app:feature_extraction_example:context}

For JavaScript/TypeScript, the pipeline identifies the enclosing functional unit and lightweight API cues.
In this case, the target line is inside a WebSocket message-receive callback (\texttt{ws.onmessage}), so we
treat the enclosing handler body as $C_\ell$.

\paragraph{Extracted evidence (illustrative)}
From $C_\ell$, we collect a small set of signals:
(i) function/event handler name (e.g., \texttt{onmessage}),
(ii) salient identifiers (e.g., \texttt{messages}, \texttt{messageDiv}, \texttt{data}, \texttt{filename}, \texttt{url}),
(iii) string literals (e.g., \texttt{"message"}, \texttt{"file"}, \texttt{"file-link"}), and
(iv) API-pattern matches indicating a WebSocket workflow (e.g., \texttt{new WebSocket} / \texttt{onmessage}).

\subsection{Step 2: Candidate Action Scoring}
\label{app:feature_extraction_example:scoring}

Let $A$ denote the taxonomy of semantic UI actions (Section~\ref{app:action}). For each candidate action $a \in A$,
we compute a relevance score $S(a, C_\ell)$ as in Eq.~\ref{app:score_func}, aggregating evidence from function names,
identifiers, string literals, route tokens (if present), and API-pattern matches, with caps on identifier/string
contributions.

\paragraph{Two competing candidates}
In this snippet, the code touches both \emph{messaging} and \emph{file sharing}. We highlight two plausible
candidates:
\begin{itemize}
    \item $\textsc{SendChatMessage}$: messaging/real-time chat interaction.
    \item $\textsc{UploadDocumentOrFile}$: file transfer functionality.
\end{itemize}

\paragraph{Score for SendChatMessage}
The WebSocket receive handler name and the presence of message-centric identifiers provide strong evidence,
augmented by an API-pattern match for WebSockets. Using the calibrated weights from Section~\ref{semantic_action_extraction} and the
identifier cap $\tau_{\text{id}}$, the contributions (illustratively) sum to:
\begin{align}
S(\textsc{SendChatMessage}, C_\ell)
&\approx w_{\text{fn}}\cdot 1
+ \min\!\bigl(\tau_{\text{id}},\, w_{\text{id}}\cdot \#\text{(message identifiers)}\bigr) \nonumber\\
&\quad + \min\!\bigl(\tau_{\text{str}},\, w_{\text{str}}\cdot \#\text{(message strings)}\bigr)
+ w_{\text{api}}\cdot 1 \nonumber\\
&= 2.5 + 1.5 + 0.35 + 1.8 = 6.15 .
\end{align}

\paragraph{Score for UploadDocumentOrFile}
Although file-related tokens appear (\texttt{file}, \texttt{filename}, \texttt{href}), they constitute weaker
support under the same weighting scheme:
\[
S(\textsc{UploadDocumentOrFile}, C_\ell)
\approx w_{\text{id}}\cdot 1.0 + w_{\text{str}}\cdot 1
= 0.5 + 0.35 = 0.85.
\]

\subsection{Step 3: Feature Assignment and Output Record}
\label{app:feature_extraction_example:assignment}

The extractor assigns
\[
\hat{a}=\arg\max_{a \in A} S(a, C_\ell) = \textsc{SendChatMessage},
\]
and outputs a compact record attached to the static-analysis finding, e.g.,
\texttt{ui\_action = send\_chat\_message} and \texttt{ui\_action\_confidence = 6.15}.
In our pipeline, this assignment is performed only when $S(\hat{a}, C_\ell)$ exceeds a small threshold
(default: $0.0$), to avoid spurious labels.

\paragraph{Interpretation}
Although computed from code for scalability in our evaluation, the extracted feature corresponds to a
\emph{black-box observable} user intent: a site with a chat interface that receives/displays messages and shared
items. This is precisely the semantic level at which FSTab operates (feature $\rightarrow$ likely backend rule IDs),
without requiring access to backend code during inference.

\paragraph{Role in the threat model}
The source-level extraction shown here is used only to label attacker-owned construction data and to make the benchmark reproducible at scale. It is not required for the victim-facing attack. At deployment time, an auditor only needs to determine whether observable actions such as chat, login, search, upload, checkout, or admin access are present something that can be done manually or with ordinary UI interaction, DOM inspection, route discovery, and network tracing. We manually verified this correspondence across the evaluated models and domains, and found that the canonicalized frontend features align with the user-visible action set recoverable in this black-box manner. The paper's contribution is the reusable feature-to-vulnerability table once those observable actions are known, not a new crawler or UI-semantic parser.

\section{End-to-End Attack Demonstration Using FSTab (Case Study)}
\label{app:attack_demo_fstab}

\paragraph{Goal}
We present an end-to-end case study showing how an attacker can leverage \textbf{FSTab} a model specific mapping from
observable UI actions to likely backend vulnerability rule IDs to prioritize and validate exploitation paths in a
black-box setting. This appendix is written for reproducibility and clarity, but follows a \emph{responsible disclosure}
style: we include measurement evidence and non-operational request skeletons, while omitting copy-pastable exploit payloads.

\paragraph{Target}
Our target is a full-stack web application (React frontend, Node/Express backend, MongoDB) generated by the \texttt{grok}
model (``Smart Matrimonial Website'') as summarized in the accompanying report.

\paragraph{Threat model}
The attacker (i) can interact with the deployed UI and send HTTP requests to public endpoints, (ii) knows the source model
identity $m$ (here: \texttt{grok}), and (iii) has access to the corresponding FSTab $T_m$, but (iv) does \emph{not} have
access to backend source code.

\subsection{From UI Reconnaissance to FSTab Query}

\paragraph{Step 1: Extract observable UI actions}
Using black-box interaction with the UI, we identify high-confidence user-intent actions such as login, quick search, and
advanced filtering. These correspond to our feature taxonomy entries and are the only inputs required to query FSTab. No source inspection is used in this step; the actions are identified from the victim's visible interface alone.

\paragraph{Step 2: Query model-specific FSTab}
Table~\ref{tab:fstab_case_query} shows the \texttt{grok}-specific top-ranked rule IDs returned for the extracted actions.
Importantly, these predictions are \emph{feature-conditioned} (e.g., search actions surface regex risks; auth actions surface
rate-limiting / injection risks), and therefore provide a concrete triage list before any code inspection.

\begin{table}[H]
\centering
\small
\begin{tabular}{l l r l}
\toprule
\textbf{UI action ($f$)} & \textbf{Predicted rule ID ($r$)} & \textbf{Score} & \textbf{Confidence} \\
\midrule
\texttt{user\_login\_with\_password} & \texttt{js/sql-injection} & 0.427 & Medium \\
\texttt{user\_login\_with\_password} & \texttt{js/regex-injection} & 0.367 & Medium \\
\texttt{apply\_search\_filter} & \texttt{js/missing-rate-limiting} & -1.507 & High \\
\texttt{search\_content} & \texttt{js/sql-injection} & 1.081 & High \\
\texttt{fetch\_data\_from\_database} & \texttt{py/sql-injection} & 1.913 & Very High \\
\bottomrule
\end{tabular}
\caption{FSTab query outputs used to guide triage in the case study (model: \texttt{grok}).}
\label{tab:fstab_case_query}
\end{table}

\paragraph{Case-specific prioritization}
Following the main-paper FSTab workflow, we prioritize (i) vulnerabilities reachable from the UI surface
(search/auth), (ii) high-impact classes (auth bypass and DoS), and (iii) those consistent with the observed backend stack
(Node/Mongo). The overall workflow is illustrated in the mapping-to-attack flow artifact.

\subsection{Validated Attacks and Measured Evidence}

\subsubsection{Attack A: Regex Injection: ReDoS (Search)}
\paragraph{Why this was selected}
Search-related UI actions strongly predicted \texttt{js/regex-injection} / \texttt{js/sql-injection} (Table~\ref{tab:fstab_case_query}),
suggesting input-to-query risks. The report identifies vulnerable use of \texttt{new RegExp(userInput)} in search routes.

\newpage 
\paragraph{Vulnerable pattern (illustrative)}
\begin{verbatim}
if (location)   query.location   = new RegExp(location, 'i');
if (occupation) query.occupation = new RegExp(occupation, 'i');
const regex = new RegExp(searchQuery, 'i');
\end{verbatim}

\paragraph{Measurement harness (non-operational)}
To demonstrate impact without providing a copy-paste exploit, we use a timing harness that measures regex processing time
as a function of adversarial input length (same structure as the provided script).
\begin{verbatim}
// Pseudocode (sanitized): measure regex evaluation time
for length in {10,15,20,25,30}:
  input = repeat("a", length) + "!"
  t_ms  = time( RegExp(<redacted_pattern>).test(input) )
  log(length, t_ms)
\end{verbatim}

\paragraph{Observed slowdown}
Table~\ref{tab:redos_numbers} reports the measured latency increase: from sub-millisecond at length 10--15 to multi-second
hangs at length 30 (DoS-scale).

\begin{table}[H]
\centering
\small
\begin{tabular}{r r l}
\toprule
\textbf{Input length} & \textbf{Time (ms)} & \textbf{Observation} \\
\midrule
10 & 0.07 & OK \\
15 & 0.24 & OK \\
20 & 6.54 & OK \\
25 & 208.26 & Slowdown \\
30 & 6655.00 & DoS-scale hang \\
\bottomrule
\end{tabular}
\caption{ReDoS evidence from the local regex test log (search feature).}
\label{tab:redos_numbers}
\end{table}

\subsubsection{Attack B: Missing Rate Limiting: Credential Stuffing / Brute Force (Auth)}
\paragraph{Why this was selected}
FSTab predicts \texttt{js/missing-rate-limiting} for multiple UI actions, and the mapping artifact highlights
auth endpoints as critical targets.

\paragraph{Validation method}
We run a rapid sequence of login requests and check whether the service returns HTTP 429 or otherwise throttles/blocks.
The provided script explicitly tests for rate limiting by breaking on 429 and logging all attempts.

\begin{verbatim}
# Pseudocode: rapid login attempts
for i in 1..N:
  code = POST /api/auth/login with fixed email
  if code == 429: stop (rate limit triggered)
  else: log attempt i and code
\end{verbatim}

\paragraph{Observed outcome}
In the recorded run, requests were not blocked/throttled (attempts proceeded without a 429), consistent with missing
rate limiting.

\subsubsection{Attack C: NoSQL Injection (Auth) -- Payload Accepted, Environment-Limited Validation}
\paragraph{Why this was selected}
FSTab prioritizes injection-class issues for auth/search actions (Table~\ref{tab:fstab_case_query}), and the analysis
points to direct use of untrusted \texttt{email} values in Mongo queries.

\paragraph{Validation notes}
We attempted operator-based injection through the login request body. In our deployment snapshot, the server returns a 500
with a MongoDB buffering timeout because the database service was not running, but the endpoint \emph{accepts} the
structured payload shape (i.e., it is not rejected by input validation), which is consistent with the predicted class.
We therefore report this as \textbf{payload accepted / validation limited by environment} (rather than a confirmed auth bypass).

\subsection{Summary and Mapping Quality (Single-Target Analogue)}

\paragraph{Attack outcomes}
Across the three attack families above, we obtain: (i) a confirmed DoS-scale slowdown in search (ReDoS),
(ii) confirmed absence of rate limiting on login attempts, and (iii) an injection attempt whose end-to-end impact is
environment-limited but consistent with the predicted vulnerability class. The executive summary and flow artifacts
summarize these results.

\paragraph{Quantitative mapping signal}
For this case study, the report summarizes an ASR of 100\% and an ACR of 50\%, meaning that the FSTab query overlapped the scanner-derived findings and covered 2/4 vulnerability types through mapping under the paper's evaluation definitions.

\paragraph{Severity (CVSS in report)}
The report assigns high-to-critical severity to the demonstrated classes (e.g., NoSQL injection critical; ReDoS high),
underscoring that FSTab-guided triage can surface practically meaningful risks early in the process. 

\paragraph{Responsible disclosure}
We intentionally omit copy-pastable exploit payload strings and endpoint-specific exploit recipes in this appendix.
Our goal is to demonstrate that \emph{FSTab reduces attacker search cost} by converting UI-observable actions into
actionable vulnerability hypotheses, validated via measurable outcomes (latency, throttling behavior, error modes),
without turning the appendix into an operational exploitation manual.
%%%%%%%%%%%%%%%%%%%%%%%%%%%%%%%%%%%%%%%%%%%%%%%%%%%%%%%%%%%%%%%%%%%%%%%%%%%%%%%
%%%%%%%%%%%%%%%%%%%%%%%%%%%%%%%%%%%%%%%%%%%%%%%%%%%%%%%%%%%%%%%%%%%%%%%%%%%%%%%

\end{document}